\documentclass[twocolumn,usenatbib]{mnras}
\usepackage{graphicx}
\usepackage{lscape}
\usepackage{bookmark}
\usepackage{threeparttablex}
\usepackage{tabularx}
\usepackage[T1]{fontenc}
\usepackage{ae,aecompl}
\usepackage{adjustbox}
\usepackage{subcaption}
\usepackage{savesym}
\usepackage{amsmath}
\usepackage{gensymb}
\savesymbol{iint}
\usepackage{wasysym}
\usepackage{graphicx}	
\usepackage{amssymb}	
\usepackage{etoolbox}
\usepackage{threeparttable}
\usepackage{environ}
\usepackage{hyperref}
\usepackage{rotating}
\usepackage{adjustbox}

\usepackage{soul}


\title[]{UOCS-XI. Study of blue straggler stars in open cluster NGC 7142 using UVIT/AstroSat}
\author[A. Panthi et al.]{
Anju Panthi,$^{1}$\thanks{p20190413@pilani.bits-pilani.ac.in}
Kaushar Vaidya,$^{1}$ 
Nagaraj Vernekar,$^{2}$ 
Annapurni Subramaniam,$^{3}$ 
\newauthor
Vikrant Jadhav, $^{4}$ 
Manan Agarwal, $^{5}$ 
\\
$^{1}$ Department of Physics, Birla Institue of Technology and Science, Pilani, Rajasthan-333031, India\\
$^{2}$ University of Padova, Via VIII Febbraio, 2, 35122 Padova PD, Italy \\
$^{3}$ Indian Institute of Astrophysics, Sarjapur Road, Koramangala, Bangalore 560034, India \\
$^{4}$ Helmholtz-Institut für Strahlen- und Kernphysik, Universität Bonn, Nussallee 14-16, D-53115 Bonn, Germany \\
$^{5}$ Anton Pannekoek Institute for Astronomy \& GRAPPA, University of Amsterdam, Science Park 904, 1098 XH Amsterdam, The Netherlands
}

\begin{document}
\label{firstpage}
\pagerange{\pageref{firstpage}--\pageref{lastpage}}
\maketitle

\begin{abstract}
We present a study of blue straggler stars (BSSs) of open cluster NGC 7142 using \textit{AstroSat}/UVIT data and other archival data. Using a machine learning-based algorithm, ML-MOC, on \textit{Gaia} DR3 data, we find 546 sources as cluster members. Based on the location on the \textit{Gaia} color-magnitude diagram, we identify ten BSS candidates, also detected in UVIT/F148W filter. We study the variable nature of BSSs by constructing their light curves using the TESS data. Two BSSs reported as eclipsing binaries in \textit{Gaia} DR3 are confirmed to be eclipsing binaries based on our analysis and also show the presence of hot companions as per the multi-wavelength spectral energy distributions (SEDs). The physical parameters of the hot companions of these two BSSs derived by fitting binary models to their light curves and those derived from the SEDs are found to be in good agreement. Additionally, five more BSSs in the cluster shows UV excess, four of which are likely to have a hot companion based on SEDs. The hot companions with the estimated temperatures $\sim$14000 $-$ 28000 K, radii $\sim$0.01 $-$ 0.05 R$_{\odot}$, and luminosities $\sim$0.03 $-$ 0.1 L$_{\odot}$, are inferred to be extremely low mass ($<$ 0.2 M$_{\odot}$), low-mass (0.2 $-$ 0.4 M$_{\odot}$), normal-mass (0.4 $-$ 0.6 M$_{\odot}$), and high-mass ($>$ 0.6 M$_{\odot}$) white dwarfs (WD). For the first time in an open cluster, we find the entire range of masses in WDs found as hot companions of BSSs. These masses imply that the  Case-A/Case-B mass transfer as well as merger are responsible for the formation of at least 60$\%$ of the BSSs of this cluster. 
\end{abstract}

\begin{keywords}
binaries: general – blue stragglers – white dwarfs – open clusters and associations: individual: NGC 7142 – ultraviolet: stars.
\end{keywords} 

\section{Introduction} \label{Section 1}

Blue straggler stars (BSSs) are the enigmatic stellar population that appears younger than other stars in their host clusters. They are identified by their location on the color-magnitude diagram (CMD), where they appear brighter and bluer than the cluster main sequence turn-off (MSTO). They were first identified in a globular cluster (GC) M3 in 1953 \citep{sandage1953color}. It is believed that BSSs can form through a variety of mechanisms, including stellar collisions \citep{hills1976stellar,chatterjee2013stellar}, mass transfer (MT; \citealt{mccrea1964extended}), and merger in hierarchical triple system \citep{perets2009triple}. The collisions can result in the formation of a single, more massive star that appears younger than the surrounding stars in the cluster. It is observed in dense stellar environments such as cores of the GCs \citep{hurley2005complete,chatterjee2013stellar,hypki2013mocca}. In the case of the MT formation channel, the BSS progenitor fills its Roche lobe and accrete mass from its companion star. This mechanism can further be classified into three sub-categories depending on the stages of the donor during MT. In Case-A MT, the donor is in the main sequence phase \citep{webbink1976evolution} while transferring the mass. On the other hand, in Case-B MT, the donor is in the red-giant branch phase during MT and eventually forms a He core WD \citep{mccrea1964extended}. Finally, in Case-C MT, the donor transfers the mass while in the asymptotic-giant branch phase \citep {chen2008binary}, and leaves behind a CO WD. When it comes to a merger, the Kozai mechanism and tidal friction may cause the formation of closely bound inner binaries in a hierarchical triple system. Over time, angular momentum loss due to either stellar evolution or a magnetized wind can cause the binaries to merge or undergo MT between each other. This process can lead to the creation of BSS in long binary or triple systems, as discussed in studies by \cite{kiseleva1998tidal}, \cite{fabrycky2007shrinking}, and \cite{naoz2014mergers}. 

Open clusters (OCs) are ideal environments for investigating the structure and history of the Galactic disk, as well as for studying the development and evolution of both single and binary star populations. The dynamical interactions among the stellar populations within a cluster can result in the formation of various exotic stellar objects, including BSSs \citep{ahumada2007new}. OCs are closer and less populated than GCs; hence, observing the properties of BSSs in OCs can help to narrow down their formation mechanisms. Moreover, by studying the BSSs population in OCs, a better understanding of the age distribution and star formation history of the host cluster can also be achieved \citep{de2006search}. The ultraviolet (UV) wavelengths are crucial in detecting BSSs because they emit a significant amount of flux within the UV range. Furthermore, the UV observations can be used to detect their hotter companions that may be present. The discovery of hot companions such as extremely low-mass (ELM; M $<$ 0.2 M$_{\odot}$) and low-mass (LM; M $<$ 0.4 M$_{\odot}$) WDs suggests that the BSSs were formed via Case-A or Case-B MT in binary systems.

The Ultraviolet Imaging Telescope (UVIT) onboard \textit{AstroSat} is capable of detecting BSSs and their hot companions when combined with other multi-wavelength data. Several studies on OCs have been conducted in the past in this context. \cite{subramaniam2016hot}, for example, reported a post AGB/horizontal-branch companion of a BSS in the OC NGC 188. \cite{sindhu2019uvit}, \cite{jadhav2019uvit}, and \cite{pandey2021uocs} investigated the OC M67 and discovered ELM WDs as BSS companions. Similarly, \cite{vaidya2022uocs} discovered ELM WDs as hot companions of BSSs in the intermediate-age OC NGC 7789. In the case of NGC 2506, two ELM WDs and one LM WD were discovered as the BSSs hot companions and normal and high-mass WDs as the hot companions of YSS \citep{panthi2022uocs}. Recently, \cite{rani2023uocs} reported A-type sub-dwarfs as hotter companions of YSS in the OC NGC 2818. 

NGC 7142 (RA = 326.287 deg, DEC = 65.775 deg) is an old OC located at the distance of $\sim$ 2 kpc \citep{sun2020wiyn} and is surrounded by molecular and dust clouds \citep{straivzys2014open}. It is present within 0.4\degree of a well-known nebula, NGC 7129, and has been the subject of several photometric and spectroscopic studies aimed at understanding the properties and evolution of different stellar populations. \citet{hoag1961johnson} presented the initial UBV photometric investigation of this cluster and showed the CMD with the red giants sequences and several stars near the MSTO. Another significant photometric study was done by \citet{van1970old}, who showed the resemblance of its CMD with the renowned OCs NGC 188 and M67. They reported the age of the cluster to be 1.5 $-$ 4 Gyr and \big[Fe/H\big] = --0.45 $\pm$ 0.2 dex, and the distance modulus $(m-M)_{V}$ ranging from 11.8 to 13.7 mag. \cite{crinklaw1991ccd} published the first CCD photometry of the cluster using the BV bands in their search for variable stars. They affirmed that variable reddening exists and suggested that its extent is less than 0.1 mag in E(B$-$V). They concluded that this cluster is older than M67 (4.0 $\pm$ 0.5 Gyr) but younger than NGC 188 (7 $\pm$ 0.5 Gyr old), and reported $(m-M)_{V}$ = 11.4 mag. Later, \cite{carraro1998galactic} compared the synthetic CMDs generated from synthetic model isochrones and reported an age of 4.9 Gyr. \cite{salaris2004age} derived an age of 4 $\pm$ 1 Gyr based on the difference in magnitude between the sub-giant branch and the red giant clump. However, \cite{janes2011quantitative} recently estimated an age of 6.9 $\pm$ 0.8 Gyr using synthetic CMDs. It is noteworthy that the differential extinction in the cluster \citep{crinklaw1991ccd, straivzys2014open} and the sparse populations of evolved stars make it challenging to locate the reference points in the CMD and may account largely for the discrepancy in the age determination of this cluster. Recently, \cite{sun2020wiyn} presented UBVRI photometry of this cluster using WIYN 0.9 m telescope and reported E (B$-$V) = 0.338 $\pm$ 0.031 mag, \big[Fe/H\big] = 0.0 $\pm$ 0.1 dex, and age = 4.0$^{-0.7}_{+1.3}$ Gyr. The initial variability study of NGC 7142 had been carried out by \cite{crinklaw1991ccd} followed by \cite{rose2007search}. The former discovered one short-period variable star near the cluster MSTO, whereas the latter found eight low-amplitude variable stars. Later, \cite{sandquist2011variable} presented the discovery of eight contact or near-contact eclipsing binaries, one long-period variable, and four detached eclipsing binaries in the cluster. 

Although NGC 7142 has been extensively researched, the BSSs in this cluster has not been subjected to detailed studies. Furthermore, the variability of the BSSs in this cluster has not been comprehensively explored. As mentioned above, NGC 7142 resembles in properties with the thoroughly examined old clusters NGC 188 and M67; therefore, it is important to explore this cluster as well. Given the importance of UV wavelengths to study BSSs, we present the UV-based analysis of BSSs of NGC 7142 for the first time. We also search for variability signatures of BSSs using Transiting Exoplanet Survey Satellite (TESS) data. Recently, a similar analysis was presented by \cite{vernekar2023photometric}, where they studied the light curves of five BSSs that are known to be SB1s in the OC M67 using Kepler and TESS data. From the light curves, they provided supporting evidence of the MT as the formation mechanism of BSSs to the earlier multi-wavelength SED analysis.

This paper is structured as follows: In \S \ref{Section 2}, the observational data and the analysis are described. The data analysis is presented in \S \ref{Section 3}. The results and discussions are given in \S \ref{Section 4}. Finally, the summary and conclusions of the work are presented in \S \ref{Section 5}.

\section{Observational data and reduction} \label{Section 2}

\subsection{Ultraviolet Imaging Telescope}

The observations of NGC 7142 were carried out using UVIT in one FUV filter, F148W, on 13$^{th}$ June 2018 under the proposal ID A04-075 with the exposure time of 2280.717 sec. UVIT is an imaging instrument onboard \textit{AstroSat}, which is made up of two Ritchey-Chrétien telescopes with 38 cm apertures. In UVIT, one telescope observes in FUV (130$-$180 nm), and the other in NUV (200$-$300 nm) and visible wavelengths (320$-$550 nm). However, the NUV channel developed a technical issue in 2018 and has not been functional since then. UVIT produces images with a Full Width Half Maximum of $\sim$1.5\arcsec and a circular field of view of 28\arcmin in diameter. There are various filters available in the FUV, NUV, and visible bands. The UV detectors operate in photon counting mode, whereas the visible detectors operate in integration mode. For the details on the effective area curves of UVIT filters and other instrument-related information, readers are referred to \cite{kumar2012ultraviolet, subramaniam2016orbit, tandon2017orbit}. We use CCDLAB \citep{postma2017ccdlab, postma2021uvit}, a customised software package, to correct for geometric distortion, flat field, spacecraft drift, astrometry and create images for each orbit. The orbit-wise images were then co-aligned and combined to generate science-ready images.

\subsection{Transiting Exoplanet Survey Satellite}

TESS is an all-sky broad-band photometric survey mission covering upto 85$\%$ of the sky in its first two years. Each hemisphere is divided into thirteen sectors, with one sector being observed for 27.4 days. It provided the data in two cadences, 2 minutes and 30 minutes, in the 2-year nominal mission, but was later augmented to 20 seconds and 200 seconds, in the extended mission. It has four identical cameras, each with a field of view of 24 $\times$ 24 degrees. Each of the four cameras is equipped with a 2K $\times$ 2K CCD, with each pixel covering 21\arcsec. The cameras are arranged so that all four can cover a 24 $\times$ 96 square degree strip of sky. The detectors are sensitive from 600 to 1000 nm, i.e., within blue to the near-IR wavelengths.

TESS light curves (LCs) of all our BSS samples are observed using different data projects. These include Quick Look Pipeline (QLP; \citealt{ huang2020photometry, kunimoto2022tess}), Cluster Difference Imaging Photometric Survey (CDIPS; \citealt{bouma2019cluster}), and a PSF-based Approach to TESS High-Quality Data Of Stellar Clusters (PATHOS; \citealt{nardiello2019psf, nardiello2021psf}). 
The QLP project generates LCs from a group of stars and other stationary luminous objects with a limited G magnitude of 13.5. For the details on the data reduction process, readers are referred to \cite{huang2020photometry}. On the other hand, the CDIPS project creates the LCs for stars that are candidate members of OCs and moving groups. Each LC corresponds to 20 to 25 days of observations of a star brighter than 16 in G magnitude. The PATHOS project generates a library of high-precision LCs for cluster members using photometry extracted from TESS Full Frame Pictures. The PSF-based technique reduces dilution effects in crowded situations, allowing for the extraction of high-precision photometry for faint stars. The PATHOS data release X has 770 LCs of 168 star clusters, including NGC 7142 \citep{messina2022gyrochronological}.
  
\section{Data analysis} \label{Section 3}

\subsection{Membership determination}
\begin{figure*}
\centering
\includegraphics[width=0.3\textwidth]{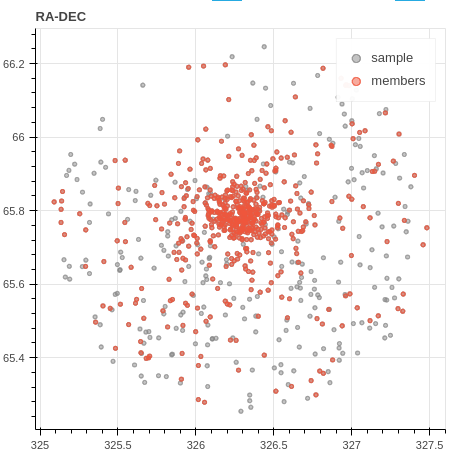}
\includegraphics[width=0.3\textwidth]{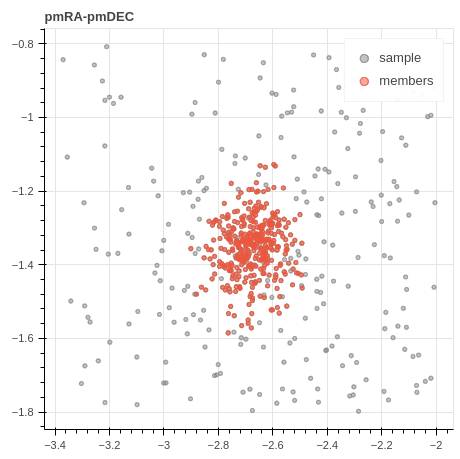}
\includegraphics[width=0.3\textwidth]{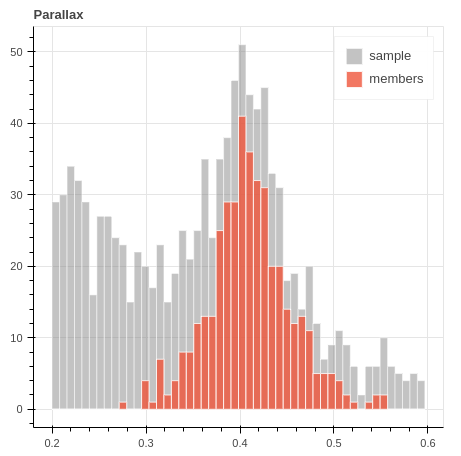}
\caption{The spatial distribution, proper motion, and parallax distributions of the members and sample sources determined using ML-MOC algorithm.}
\label{Fig.1}
\end{figure*}

In the case of OCs, the contamination due to field stars is a big challenge. Hence, obtaining secure membership is of utmost importance to do any further analysis. We used ML-MOC algorithm \citep{agarwal2021ml} on \textit{Gaia} DR3 \citep{vallenari2022gaia} data in order to determine the cluster members. This is a machine-learning-based algorithm that does not require any prior knowledge of a cluster and employs the k-Nearest Neighbour (kNN, \citealt{cover1967nearest}) and the Gaussian mixture model (GMM, \citealt{peel2000robust}) algorithms. The steps that we followed to determine cluster members using ML-MOC are described briefly here. First, we downloaded the sources within 30$\arcmin$ radius of the cluster center that have five astrometric parameters (positions, proper motions, and parallax), appropriate measurements in three Gaia photometric passbands G, G$_{BP}$, and G$_{BP}$, non-negative parallaxes, and error in G-mag less than 0.005. These are called \textit{All sources}. Then, we estimated the mean proper motions and mean parallaxes of the cluster using the kNN algorithm to determine the ranges of proper motions and parallaxes that enclose all of the likely cluster members. These are termed as \textit{Sample sources}. Next, we apply the GMM, an unsupervised clustering algorithm, to the \textit{Sample sources} in order to separate the likely cluster members from the field stars by fitting two Gaussian distributions in the proper motions and parallax space of the \textit{Sample sources}. The fitting of GMM is done by maximizing the likelihood of the distribution parameters using the expectation maximization algorithm \citep{dempster1977maximum}. The K-component GMM, for N data points in an M-dimensional parameter space, is defined as 
\begin{equation}
P(x) = \sum_{i=1}^{K} \omega_{i} G(x| \mu_{i}, \Sigma_{i})
\end{equation}
%such that $\sum_{i=1}^{K} \omega_{i}$ = 1
where P(x) represents the probability distribution of data points x and $\omega_{i}$ is the mixture weight of the i$^{th}$ Gaussian component, G(x| $\mu_{i}, \Sigma_{i}$). This Gaussian component is
defined as 
\begin{equation}
G(x| \mu_{i}, \Sigma_{i}) = \frac{\exp\left[-\frac{1}{2}(x - \mu_i)^T \Sigma_i^{-1} (x - \mu_i)\right]}{{(2\pi)^{M/2}\sqrt{|\Sigma_i|}}}
\end{equation}
In both the above equations, $\mu_{i}$ is the mean vector, and $\Sigma_{i}$ is the full covariance matrix of the i$^{th}$ Gaussian component. The GMM calculates a probability that designates the likelihood of each data point belonging to the cluster. We use a two-component, i.e., cluster and field, GMM on the normalized three-dimensional parameter space ($\mu \alpha$, $\mu \delta$, $\omega$) assuming that the proper motions and parallaxes of the \textit{Sample sources} follow a two-component Gaussian distribution. Our member selection criteria considers sources with high membership probabilities exceeding 0.6 as members. Subsequently, we expand this list to include \textit{Sample sources} whose parallax values lie within the range of parallaxes of even higher probability members ($>$ 0.8). The members added later are low-probability members with probabilities ranging between 0.2 and 0.6. Out of the 546 sources identified as cluster members, only six are with probabilities ranging between 0.2 and 0.6; the rest are high probability members. The proper motion, spatial, and parallax distributions of sample sources and the identified cluster members are shown in Figure \ref{Fig.1}.  
  
\subsection{Photometry and Color-magnitude diagram}

We used the DAOPHOT package of IRAF \citep{stetson1987daophot} in order to perform the point-spread function (PSF) photometry and obtained the magnitudes of the detected stars in FUV image, F148W. To accomplish this task, we employed the following steps. First, we estimated the background counts and mean FWHM of the image using \textit{imexamine} task. Then, we detected the sources above a certain threshold limit using \textit{daofind} task in IRAF. We performed the aperture photometry of the sources using the \textit{phot} task. In order to perform PSF photometry, we selected the bright and isolated sources using \textit{pstselect} and created a PSF model using \textit{psf} task. Finally, PSF magnitudes of all the stars were estimated simultaneously using multiple iterations of PSF fitting to the group of stars using \textit{allstar} task. We also applied the PSF corrections in order to eliminate the systematic difference between \textit{allstar} magnitudes and \textit{phot} magnitudes, followed by the aperture correction, which was determined using the curve-of-growth technique. Moreover, since the FUV detector of UVIT works in photon counting mode, it may be subjected to counts more than one photon per frame. In order to eliminate this issue, we applied the saturation corrections following \cite{tandon2020additional}. \\

\begin{figure*}
\centering
\includegraphics[width=0.4\textwidth]{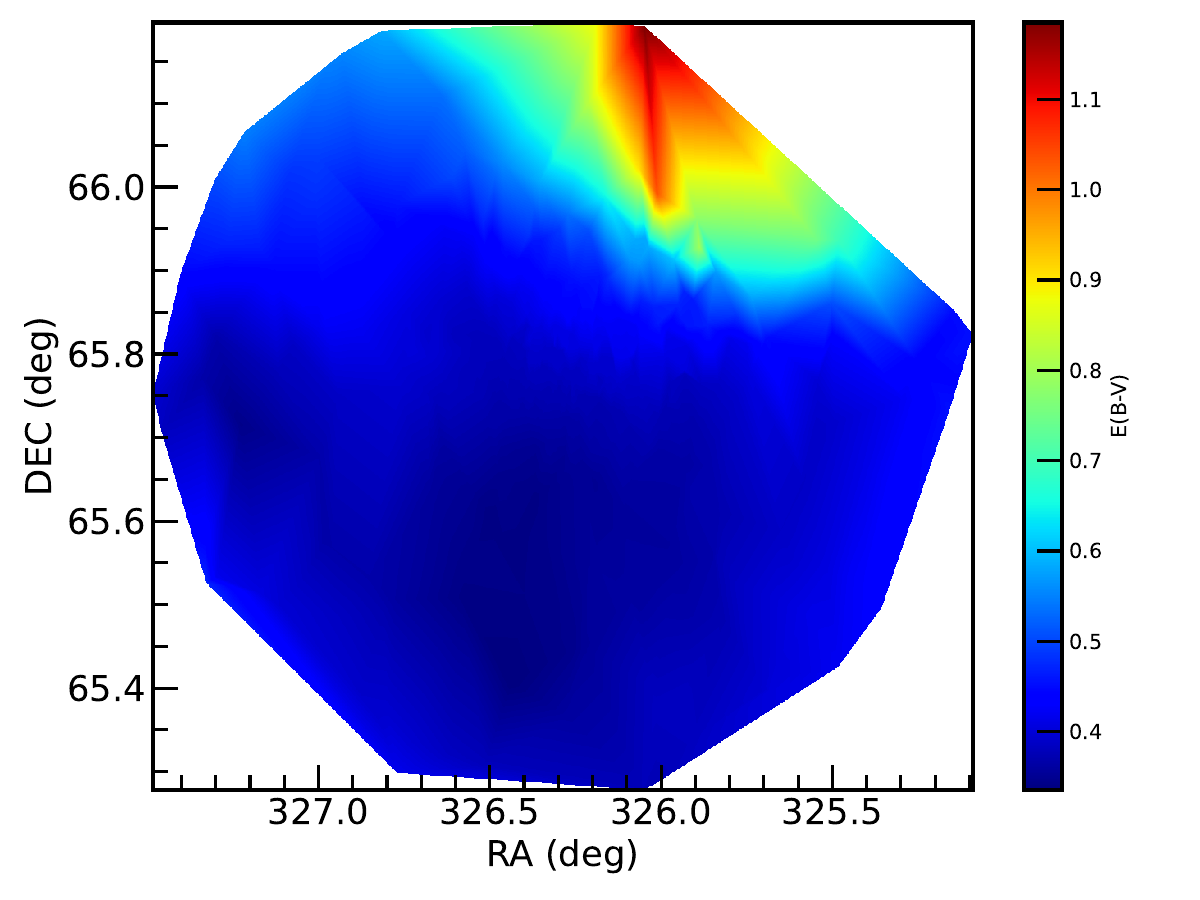}
\includegraphics[width=0.4\textwidth]{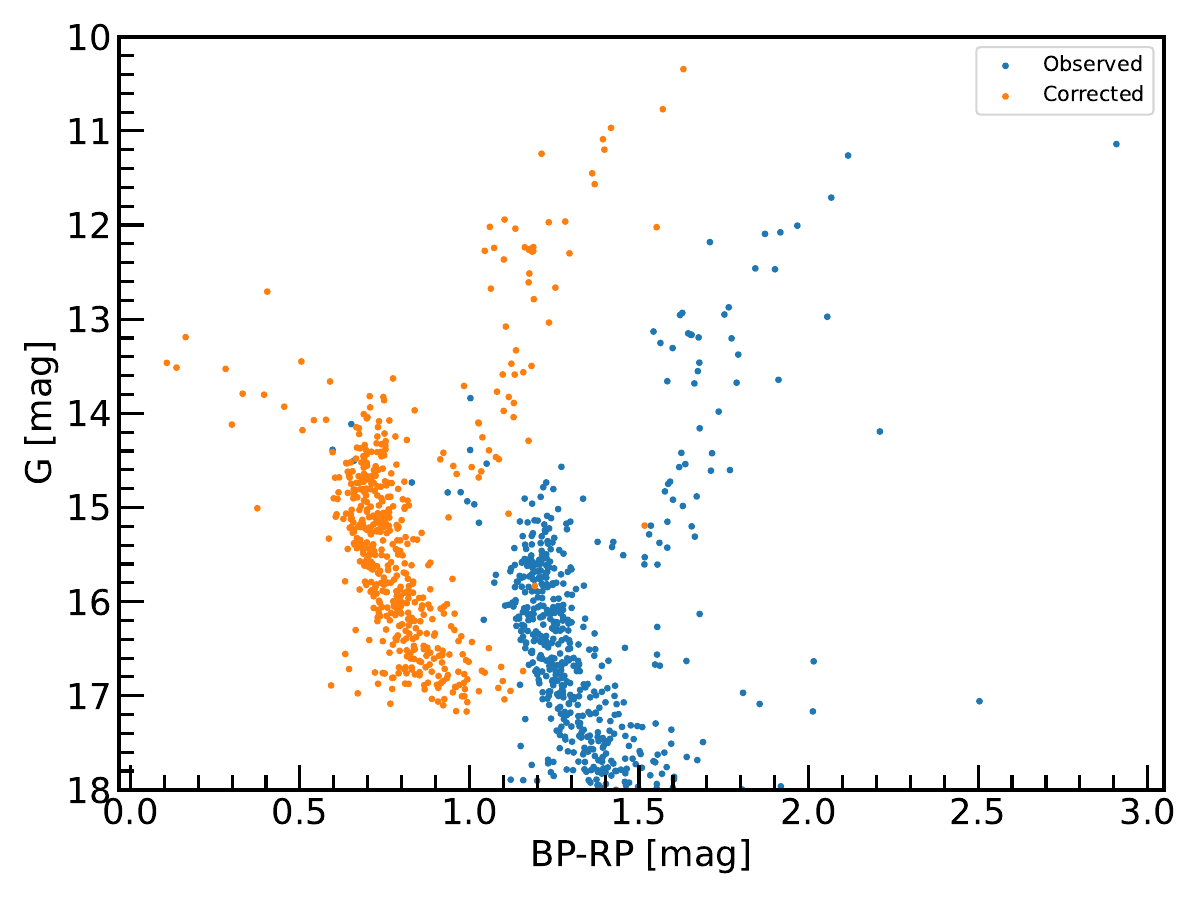}
\caption{The left panel shows the reddening map of the cluster members. The right panel shows the observed \textit{Gaia} DR3 CMD as blue dots and differential reddening and extinction corrected CMD as orange dots.}

\label{Fig.2}
\end{figure*}

To quantify the differential reddening and correct it, we used the method adopted in \cite{della2023ongoing}. First, we select the likely main sequence stars (G $>$ 15 mag or G$_{BP}$ $-$ G$_{RP}$ $<$ 1.5 mag) from the members determined using ML-MOC. Then we find counterparts of these members in Pan-STARRS \citep{chambers2016pan} photometric bands \textit{r}, \textit{i}, and \textit{z} by searching within 1\arcsec of the Gaia coordinates. Next, we plot a color-color diagram, \textit{G$-$r} vs \textit{i$-$z}. This diagram is a suitable choice as the evolutionary sequences are generally orthogonal to the reddening vector in these colors \citep{della2023ongoing}. We refer to the median of this color-color distribution as our reference point. Afterward we obtain the median colors of the closest 50 stars of each main sequence star and estimated the distance of this median value to the reference system along the reddening vector using the coefficients given in \cite{cardelli1989relationship}. For the remaining stars which are not in the main-sequence, we assign the median reddening of the closest 50 neighbors \citep{della2023ongoing}. At last, we apply the extinction value corresponding to the derived distance to member stars and plotted the reddening map, as shown in the left panel of Figure \ref{Fig.2}. It is noteworthy that the reddening is more severe in the northwest field; however in the center as well as in the southern regions, the reddening is low and does not vary significantly. The significantly high reddening in the northwest direction can be attributed to the presence of the NGC 7129 nebula in the same direction. The right panel of Figure \ref{Fig.2} shows the optical CMD of the cluster members determined using ML-MOC before and after the differential reddening and extinction corrections.

\begin{figure}
\centering
\includegraphics[width=\columnwidth]{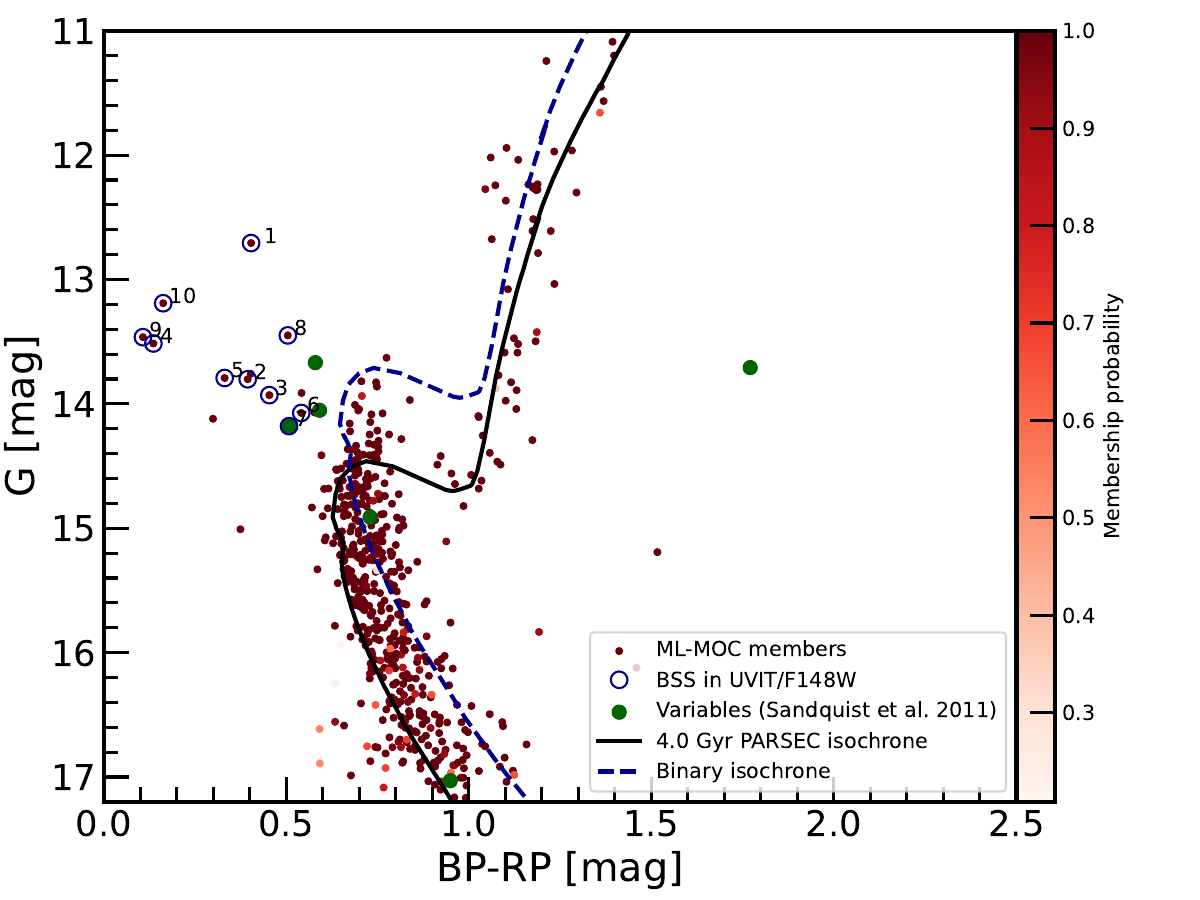}
\caption{The differential reddening and extinction corrected optical CMD of NGC 7142 showing members determined using ML-MOC, BSSs detected in UVIT/F148W filter, a PARSEC isochrone, and an equal mass binary sequence.}
\label{Fig.3}
\end{figure}

Figure \ref{Fig.3} shows the differential reddening and extinction corrected optical CMD of the cluster. The cluster members determined using the machine-learning-based algorithm, ML-MOC, are shown in red dots along with their membership probabilities. The 10 BSSs identified as members and with counterparts in the UVIT/F148W are shown as blue open circles, labeled according to the coordinates given in Table \ref{Table1}. There are four additional BSSs candidates which are cluster members but have no UV counterparts. We only study the 10 BSSs, which are members of the cluster and detected in the UVIT filter in this work. It is noteworthy that all the BSSs are highly probable cluster members with a membership probability of $\sim$0.8 - 0.9. Furthermore, the variable stars identified in a study by \cite{sandquist2011variable}, which are cluster members as per our criteria, are shown as green-filled circles. BSS 7 is one of the five variable stars from their study. The remaining four variables do not show variable nature using \textit{Gaia} or TESS data. A PARSEC isochrone \citep{bressan2012parsec} of age = 4.0 Gyr, distance = 2368 pc, and \big[Fe/H\big] = 0 \citep{sun2020wiyn} is plotted after applying the extinction correction of A$_{\text{G}}$ = 0.09 and reddening of E(B$_{\text{P}}$ $-$ R$_{\text{P}}$) = 0.03. Furthermore, we have also shown a binary sequence isochrone as the blue dashed curve for equal-mass binaries that have a G-band magnitude brighter than 0.75 mag than the MS stars. It is evident from the CMD that G-band magnitudes of BSSs range from $\sim$0.4 to $\sim$1.4 above the MSTO.

\begin{table*}
\centering
\caption{Coordinates of all BSSs in Columns 2 and 3, UVIT F148W in Columns 4, GALEX NUV flux in Column 5, \textit{Gaia} DR3 Gbp, G, and Grp fluxes in Columns 6--8, 2MASS J, H, and Ks fluxes in Columns 9--11, and WISE W1, W2, W3, and W4 fluxes in Columns 12--15. All flux values are extinction corrected and listed in the unit of erg s$^{-1}$ cm$^{-2}$\AA$^{-1}$.}

\begin{tabular}{cccccccccccc}
\hline
\hline
Name&RA&DEC&UVIT.F148W$\pm$err&GALEX.NUV$\pm$err&\\
PS1.g$\pm$err&GAIA3.Gbp$\pm$err&GAIA3.G$\pm$err&PS1.r$\pm$err&PS1.i$\pm$err&\\
GAIA3.Grp$\pm$err&PS1.z$\pm$err&PS1.y$\pm$err&2MASS.J$\pm$err&2MASS.H$\pm$err&\\
2MASS.Ks$\pm$err&WISE.W1$\pm$err&WISE.W2$\pm$err&WISE.W3$\pm$err&WISE.W4$\pm$err
\\
\hline
\hline
\\
BSS 1&325.9511&65.8598&2.361e-17$\pm$7.643e-21&3.631e-16$\pm$2.728e-17&\\
8.675e-15$\pm$4.939e-17&8.129e-15$\pm$3.039e-17&7.291e-15$\pm$1.941e-17&8.142e-15$\pm$1.010e-17&6.525e-15$\pm$2.584e-16&\\
6.370e-15$\pm$2.704e-17&5.725e-15$\pm$2.515e-17&5.114$\pm$1.917e-17&3.204e-15$\pm$7.969e-17&1.546e-15$\pm$4.273e-17&\\
6.225e-16$\pm$1.089e-17&1.262e-16$\pm$2.673e-18&3.672e-17$\pm$7.441e-19&5.706e-19$\pm$1.934e-19&7.371e-19$\pm$2.748e-19\\
\hline
\\
BSS 2&325.9342&65.8757&1.679e-17$\pm$5.301e-21&1.141e-16$\pm$2.201e-17&\\3.188e-15$\pm$3.971e-18&
2.975e-15$\pm$8.663e-18&2.659e-15$\pm$6.804e-18&3.016e-15$\pm$8.711e-18&2.477e-15$\pm$1.662e-17&\\
2.311e-15$\pm$8.389e-18&2.062e-15$\pm$5.176e-18&1.838e-15$\pm$1.327e-18&1.156e-15$\pm$2.875e-17&4.806e-16$\pm$1.682e-17&\\2.076e-16$\pm$7.459e-18&4.288e-17$\pm$9.479e-19&1.213e-17$\pm$2.906e-19&3.508e-19$\pm$1.460e-19&1.424e-18$\pm$6.521e-19\\
\hline
\\
BSS 3&326.3274&65.7773&2.365e-17$\pm$7.672e-21&1.627e-16$\pm$2.349e-17\\
3.662e-15$\pm$6.643e-18& 3.332e-15$\pm$9.242e-18&2.899e-15$\pm$7.439e-18&3.306e-15      9.287e-18&2.644e-15$\pm$1.322e-17\\
2.454e-15$\pm$9.100e-18&2.161e-15$\pm$8.379e-18&1.920e-15$\pm$1.180e-17&1.128e-15      3.846e-17&5.150e-16$\pm$1.707e-17\\
2.123e-16$\pm$7.625e-18&4.304e-17$\pm$9.117e-19&1.238e-17$\pm$2.737e-19&5.034e-19$\pm$      7.996e-20&1.205e-18$\pm$1.914e-19\\
\hline
\\
BSS 5&326.1820&65.7889&2.463e-17$\pm$9.270e-21&3.198e-16$\pm$4.833e-17\\
4.229e-15$\pm$4.538e-18&3.836e-15$\pm$1.073e-17&3.196e-15$\pm$8.152e-18&3.637e-15$\pm$8.130e-18&2.756e-15$\pm$7.566e-18\\
2.563e-15$\pm$9.162e-18&2.184e-15$\pm$3.453e-18&1.920e-15$\pm$4.722e-18&1.032e-15$\pm$3.327e-17&4.775e-16$\pm$1.363e-17\\
1.988e-16$\pm$6.410e-18&3.766e-17$\pm$7.977e-19&1.025e-17$\pm$2.455e-19&4.969e-19$\pm$    8.260e-20&7.833e-19$\pm$1.302e-19\\
\hline
\\
BSS 7&326.3131&65.823&3.050e-17$\pm$1.509e-20&2.531e-16$\pm$5.967e-17\\
2.600e-15$\pm$8.428e-18&2.396e-15$\pm$1.831e-17&2.155e-15$\pm$6.738e-18&2.490e-15$\pm$3.566e-18&2.085e-15$\pm$3.831e-18\\
1.922e-15$\pm$1.337e-17&1.795e-15$\pm$1.936e-18 &1.517e-15$\pm$1.203e-17&8.908e-16$\pm$2.379e-17&4.252e-16$\pm$1.409e-17\\
1.821e-16$\pm$6.209e-18&4.005e-17$\pm$8.854e-19&1.105e-17$\pm$2.546e-19&6.942e-19$\pm$1.800e-19&6.809e-19$\pm$1.766e-19 \\
\hline
\\
BSS 8&326.3476&65.80431&2.387e-17$\pm$8.150e-21&2.745e-16$\pm$2.814e-17&\\
5.206e-15$\pm$8.971e-18&4.886e-15$\pm$1.423e-17&4.391e-15$\pm$1.122e-17&4.974e-15$\pm$8.664e-18&4.130e-15$\pm$2.978e-18\\
3.826e-15$\pm$1.364e-17&3.415e-15$\pm$3.736e-18&3.044e-15$\pm$2.063e-17&1.888e-15$\pm$5.045e-17&8.602e-16$\pm$3.169e-17\\
3.433e-16$\pm$1.043e-17&7.249e-17$\pm$1.735e-18&2.051e-17$\pm$4.913e-19&8.927e-19$\pm$1.006e-19&1.240e-18$\pm$1.398e-19\\
\hline
\label{Table1}
\end{tabular}
\end{table*}

\subsection{Light curve analysis}

\begin{figure*}
\centering
\includegraphics[width=0.5\textwidth]{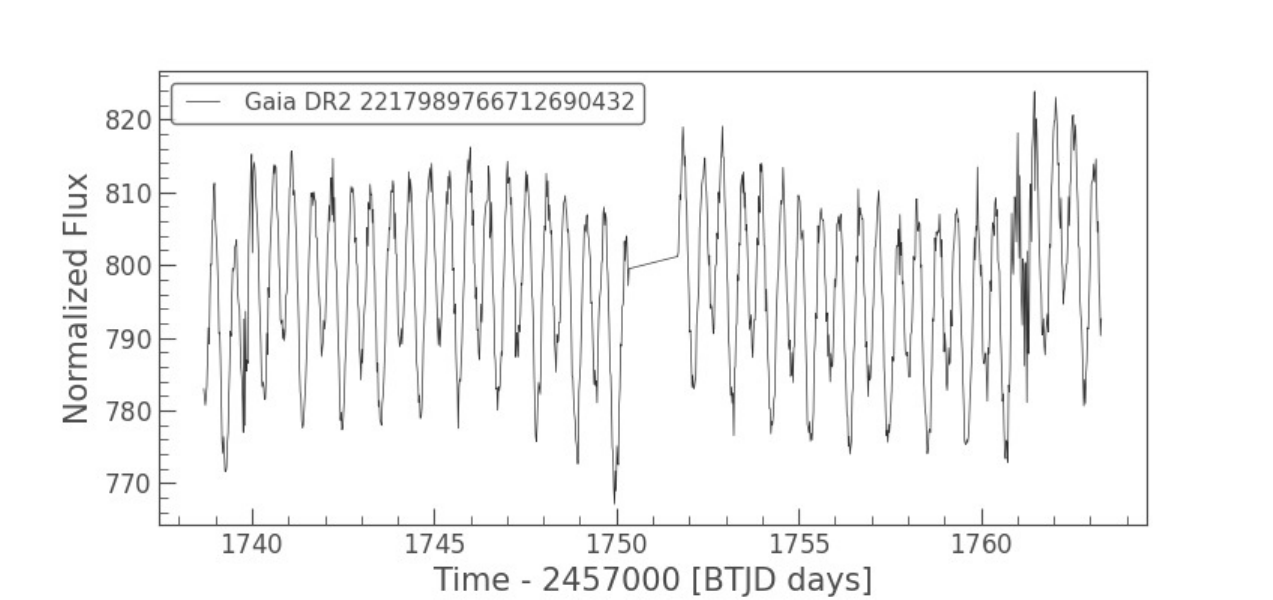}
\includegraphics[width=0.5\textwidth]{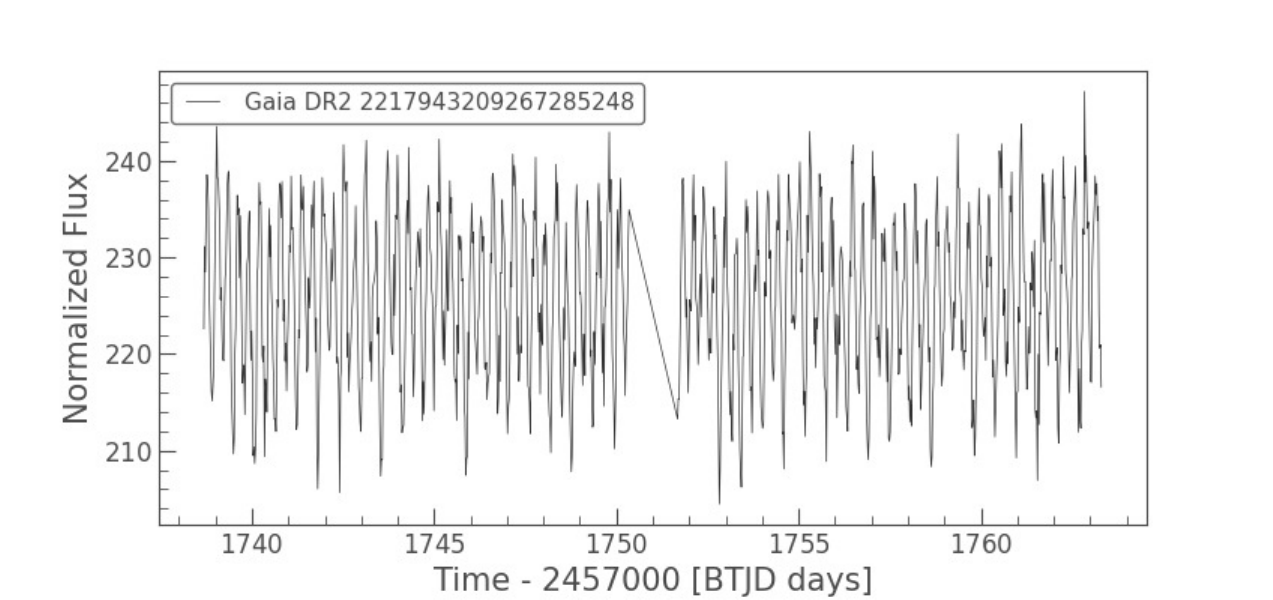}
\caption{The PATHOS lightcurve of BSS 1 and BSS 7 respectively.}
\label{Fig.4}
\end{figure*}

\begin{figure*}
\centering
\includegraphics[width=0.5\textwidth]{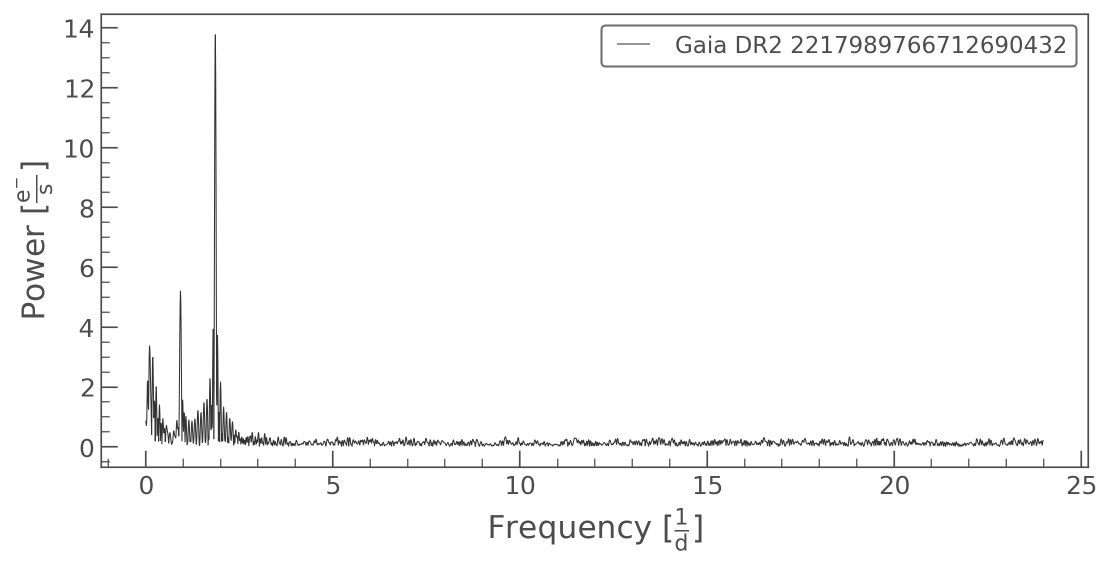}
\includegraphics[width=0.5\textwidth]{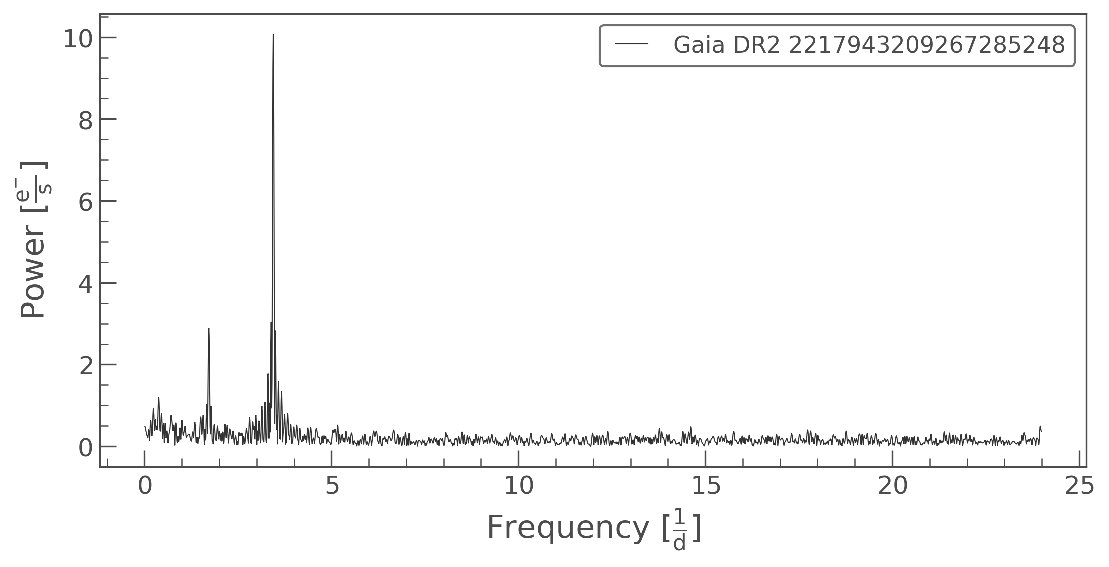}
\caption{The PATHOS periodograms of BSS 1 and BSS 7. For each of them, the first peak represents the orbital period, whereas the second peak represents the first harmonics. %The orbital period of BSS 1 is $\sim$1 day and the orbital period of BSS 7 is $\sim$0.5 days.
}
\label{Fig.5}
\end{figure*}

\begin{figure*}
\centering
\includegraphics[width=0.5\textwidth]{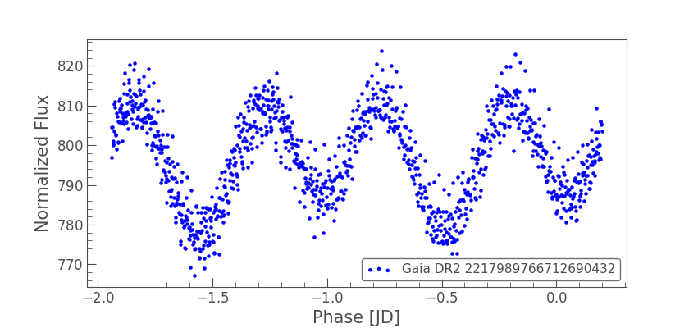}
\includegraphics[width=0.5\textwidth]{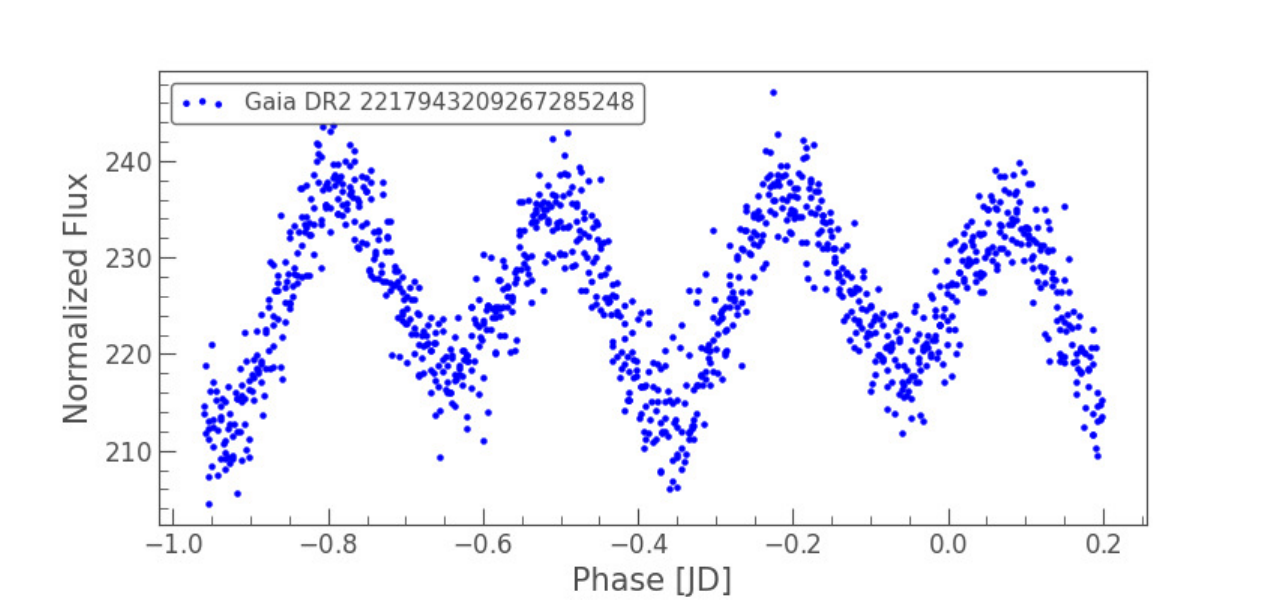}
\caption{Phase folded LCs of BSS 1 and BSS 7. The LCs are folded using the period determined in the periodograms shown in Figure \ref{Fig.5}.}
\label{Fig.6}
\end{figure*}

Variable stars can provide a wealth of new information, allowing us to confirm current theories that seek to predict cluster stellar properties from basic principles \citep{sandquist2011variable}. For example, pulsating stars can reveal information about stellar interior structure and mass. Short-period semi-detached and contact binaries may carry encoded information about the dynamical history of the cluster. Detached eclipsing binaries can be particularly valuable since they can offer high-precision measurements of masses and radii for individual stars with minimal theoretical interpretation. 
In order to check if any of the BSS is variable in nature, we used the TESS \citep{ricker2015transiting} data. We searched for the TESS LCs of all our BSS samples in different data releases, as mentioned above. In this work, we have presented the LCs of PATHOS data release. We used \textit{Lightkurve} \citep{2018ascl.soft12013L}, a python based package, to construct and analyse the LC of the BSSs. We found that two of our ten BSSs candidates (BSS 1 and BSS 7) showed some signatures of variability in sectors 16, 18, and 24, using all the three pipelines mentioned above. The LCs of these BSSs of sector 16, generated using the PATHOS pipeline, are shown in Figure \ref{Fig.4}. Then, we computed the power spectrum using Fourier transform (FT) from the LC in order to determine the period as shown in Figure \ref{Fig.5}. We take all the significant frequencies from the periodogram. A frequency is considered significant if the amplitude or power of that frequency is four times the noise \citep{breger1993nonradial}. After doing it for all the frequencies, the orbital frequency will give a phase folded curve with two dips resembling two eclipses, whereas all others will give sinusoidal variations as shown in Figure \ref{Fig.6}. It can be noted that the orbital period of BSS 1 is $\sim$1 day, whereas the period of BSS 7 is $\sim$0.5 days. \cite{sandquist2011variable} categorised BSS 7 as an eclipsing contact or near-contact binary with an orbital period of $\sim$ 0.58 days, using high -resolution spectrograph at Hobby-Eberly Telescope. However, none of their other variables were found to be variable in our analysis. 

\section{Results and Discussion} \label{Section 4}

\subsection{Spectral energy distributions}

Stars emit electromagnetic radiation across the entire frequency or wavelength range. The Spectral Energy Distribution (SED) describes the distribution of this radiation over wavelengths. Analyzing the SEDs of stars allows us to estimate their fundamental parameters and thus gain a better understanding of their formation and evolution. Furthermore, the excess in UV fluxes can indicate the likelihood of a hotter companion of BSSs, hence constraining their formation mechanism. We used the virtual observatory SED analyser (VOSA, \citealt{bayo2008vosa}) to create the SEDs. We found that three of the ten BSS candidates (BSS 4, BSS 9, and BSS 10) in this cluster have a nearby source within 3\arcsec, and hence we did not construct their SEDs. 

We obtained photometric fluxes of sources in NUV from GALEX \citep{martin2005galaxy}, optical from \textit{Gaia} DR3 \citep{babusiaux2022gaia} and PAN-STARRS \citep{chambers2016pan}, near-IR from Two Micron All Sky Survey (2MASS, \citealt{cohen2003spectral}), and far-IR from Wide-field Infrared Survey (WISE, \citealt {wright2010wide}), using VOSA. The photometric fluxes were corrected for extinction by VOSA according to the extinction law by \cite{fitzpatrick1999correcting} and \cite{indebetouw2005wavelength} using the value of extinction A$_{v}$ = 1.04$\pm$0.09 provided by us, which is taken from \cite{sun2020wiyn}. Table \ref{Table1} shows the values of the BSS fluxes in various filters.

\begin{figure}
\includegraphics[width=\columnwidth]{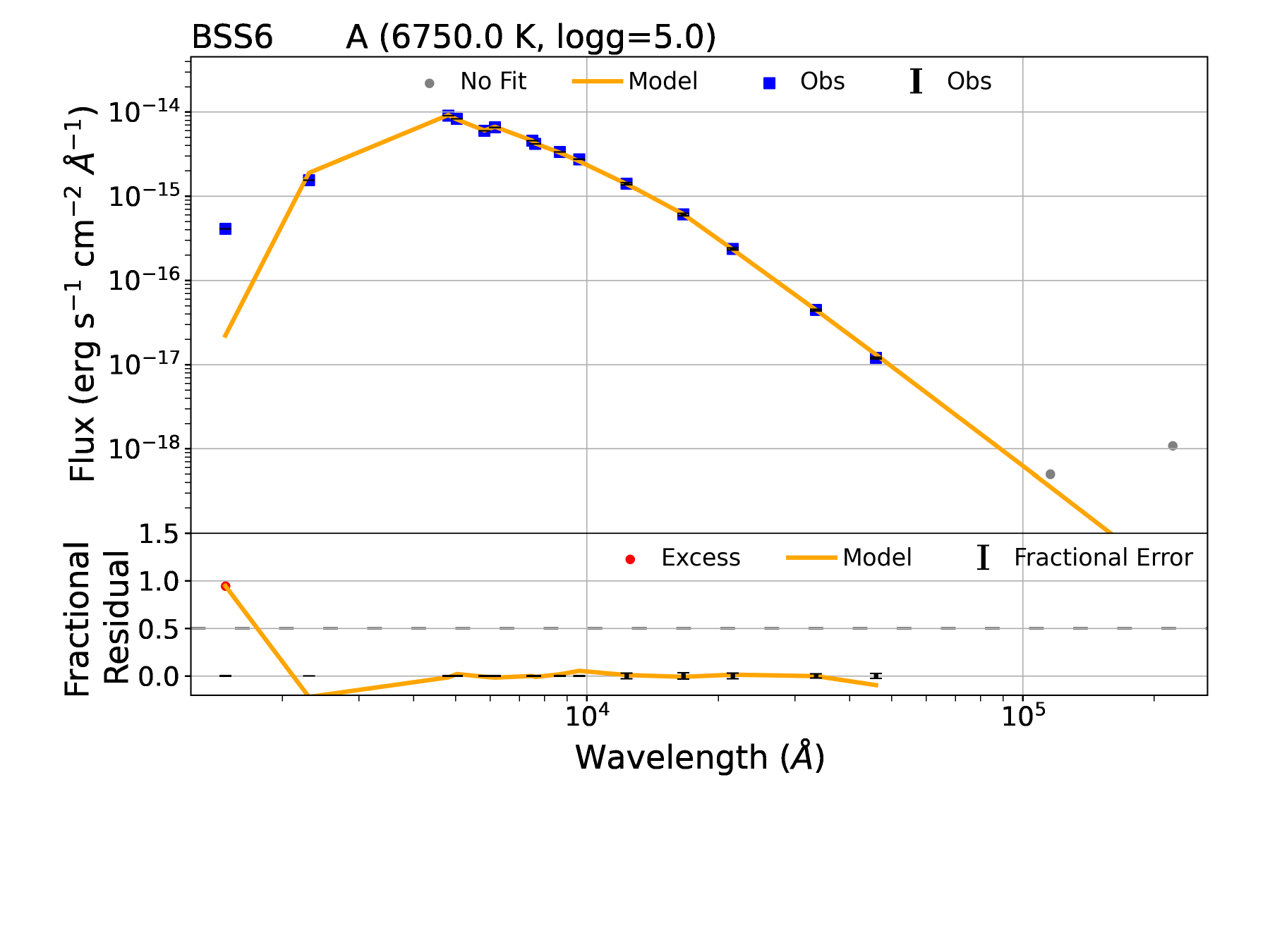}
\caption{Single-component SED fitting of BSS 6 showing UV excess. In the top panel, blue data points show the extinction corrected observed fluxes, where black error bars represents the errors in the observed fluxes, and the orange curve represents the Kurucz stellar model fit. The bottom panel shows the residual between extinction-corrected observed fluxes and the model fluxes across the filters from UV to IR wavelengths.}
\label{Fig.7}
\end{figure}

VOSA calculates synthetic photometry for selected theoretical models using filter transmission curves and performs a $\chi^{2}$ minimization test by comparing the synthetic photometry with the extinction-corrected observed fluxes to determine the best-fitting SED parameters. The following formula is used to calculate the reduced $\chi^{2}_{r}$: 

\begin{equation} 
   \chi_{r}^{2} =\frac{1}{N-N_{f}}\sum_{i=1}^{N} \frac{(F_{o,i}-M_{d}F_{m,i})^{2}}{\sigma^{2}_{o,i}}
\end{equation} 

where N is the number of photometric data points used and N$_{f}$ is the number of model-free parameters. The observed and model fluxes of the star are denoted by F$_{o,i}$ and F$_{m,i}$ respectively. M$_{d}$ is the scaling factor that must be multiplied by the model to obtain the fit, and it is given by (R/D)$^{2}$, where R is the radius of the star, D is the distance to the star, and $\sigma_{o,i}$ is the error in the observed flux. We used Kurucz stellar models \citep{castelli1997notes} to fit the SEDs to the BSSs. We kept T$_{eff}$ and log g as free parameters, with ranges of 3500 -- 50000 K and 3 -- 5, respectively. We set the metallicity (\big[Fe/H\big]) value to zero, which is the cluster metallicity reported by \cite{sun2020wiyn}. We first removed the UV data points from the SEDs and checked whether the optical and IR data points fit satisfactorily with the model flux. We carefully examined each UV filter to see if there was any excess to the UV and/or IR data points. We consider sources with UV excess greater than 50$\%$ with respect to the single component model for the binary component SED fits. It is interesting to note that all the seven BSSs showed the excess in UV fluxes greater than 50$\%$, and therefore, we attempted the binary-component SED fits to them. Since we select stars based on a 50$\%$ flux excess, it is not surprising that we find that all candidates have excess larger than this threshold. BSS 6 could not be fitted with the binary component SED since all the models fitting the data points showed the highest temperatures and hence were unreliable. The single-component SED of BSS 6 is shown in Figure \ref{Fig.7}.

\begin{figure*}
    \centering
    \begin{subfigure}[b]{0.48\textwidth}
        \includegraphics[width=0.8\textwidth]{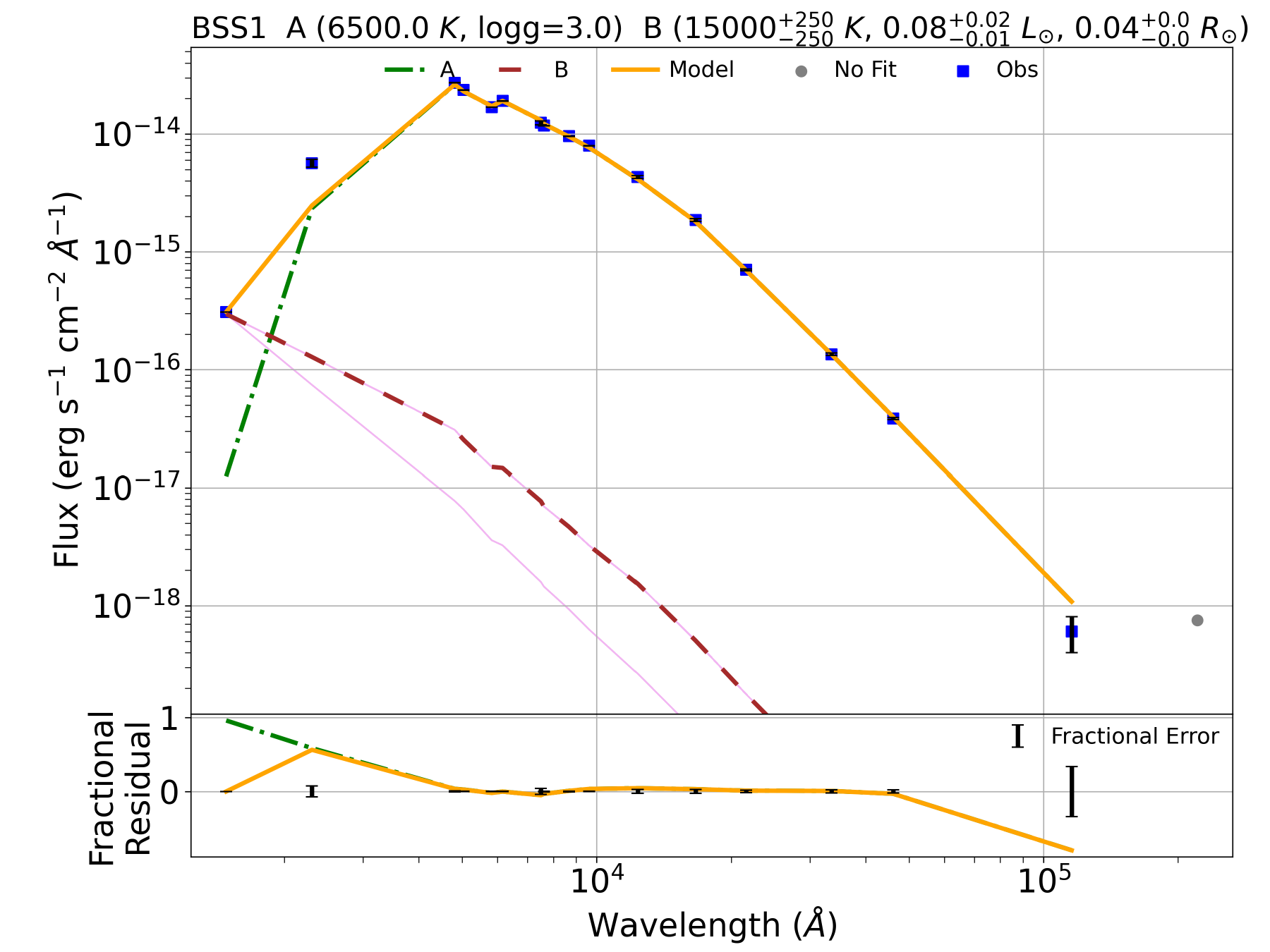}
        \caption*{}
    \end{subfigure}
    \quad
    \begin{subfigure}[b]{0.48\textwidth}
        \includegraphics[width=0.8\textwidth]{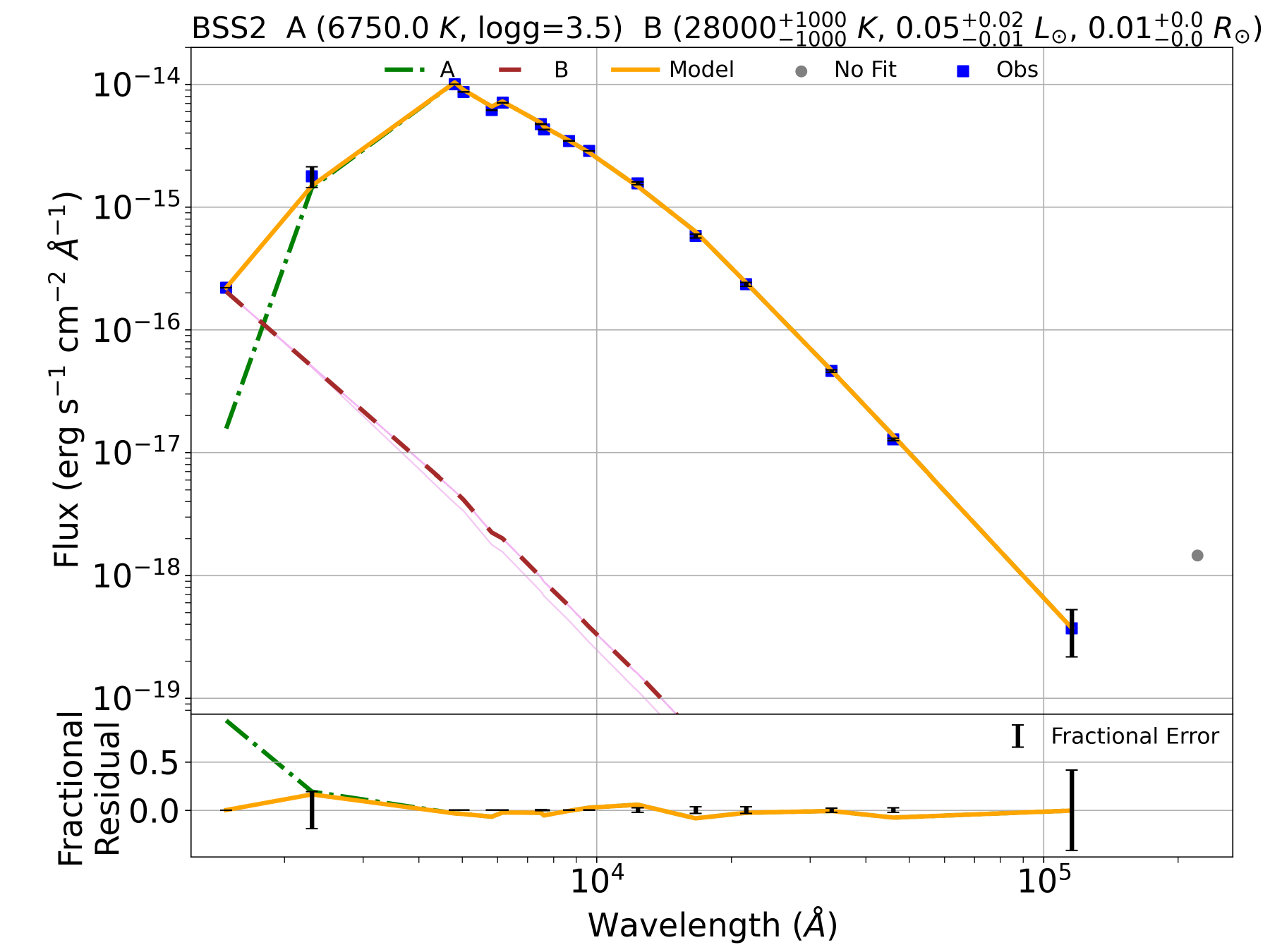}
        \caption*{}
    \end{subfigure}
    \quad
    \begin{subfigure}[b]{0.48\textwidth}
        \includegraphics[width=0.8\textwidth]{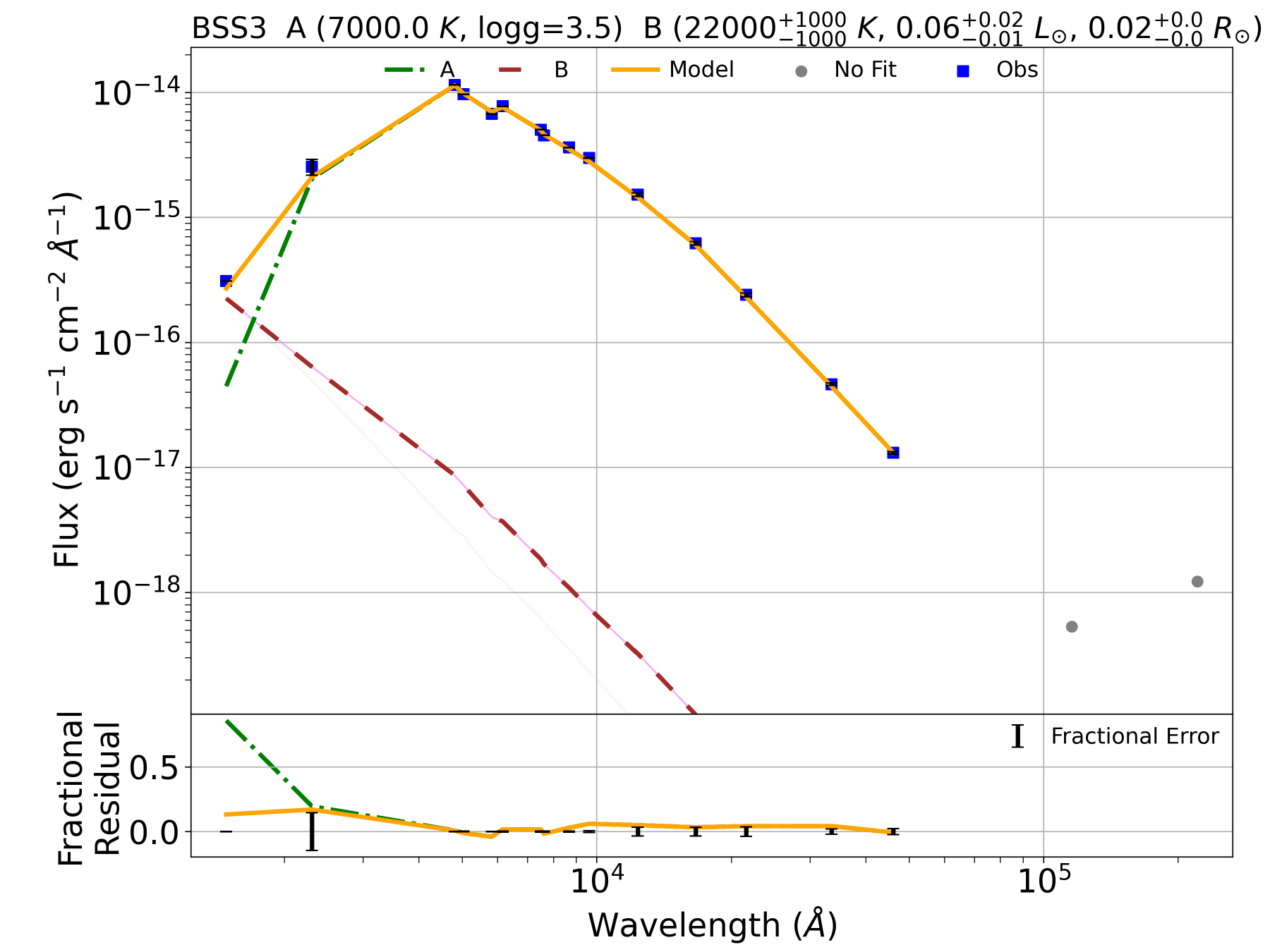}
        \caption*{}
    \end{subfigure}
    \begin{subfigure}[b]{0.48\textwidth}
        \includegraphics[width=0.8\textwidth]{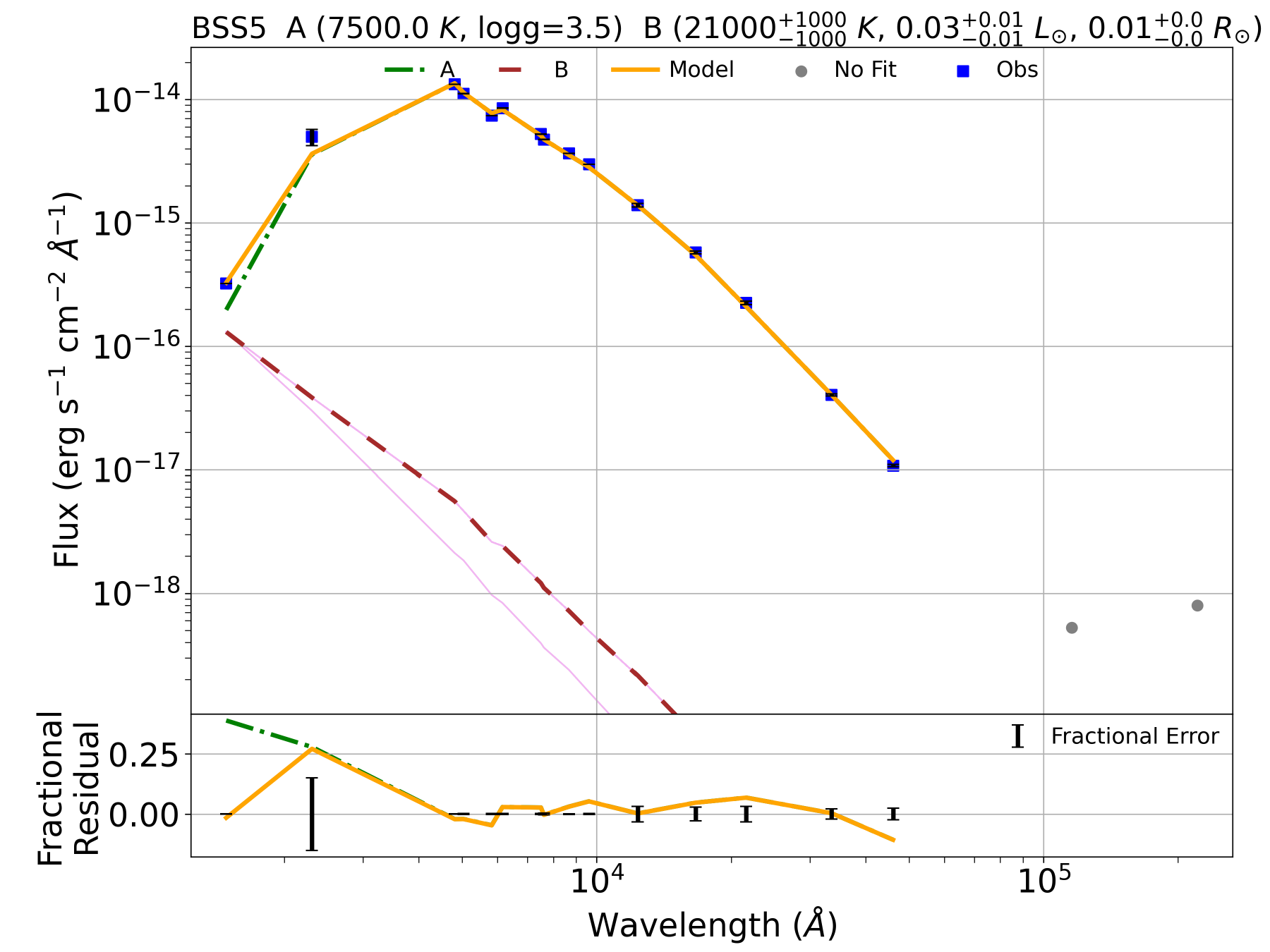}
        \caption*{}
    \end{subfigure}
    \begin{subfigure}[b]{0.48\textwidth}
        \includegraphics[width=0.8\textwidth]
{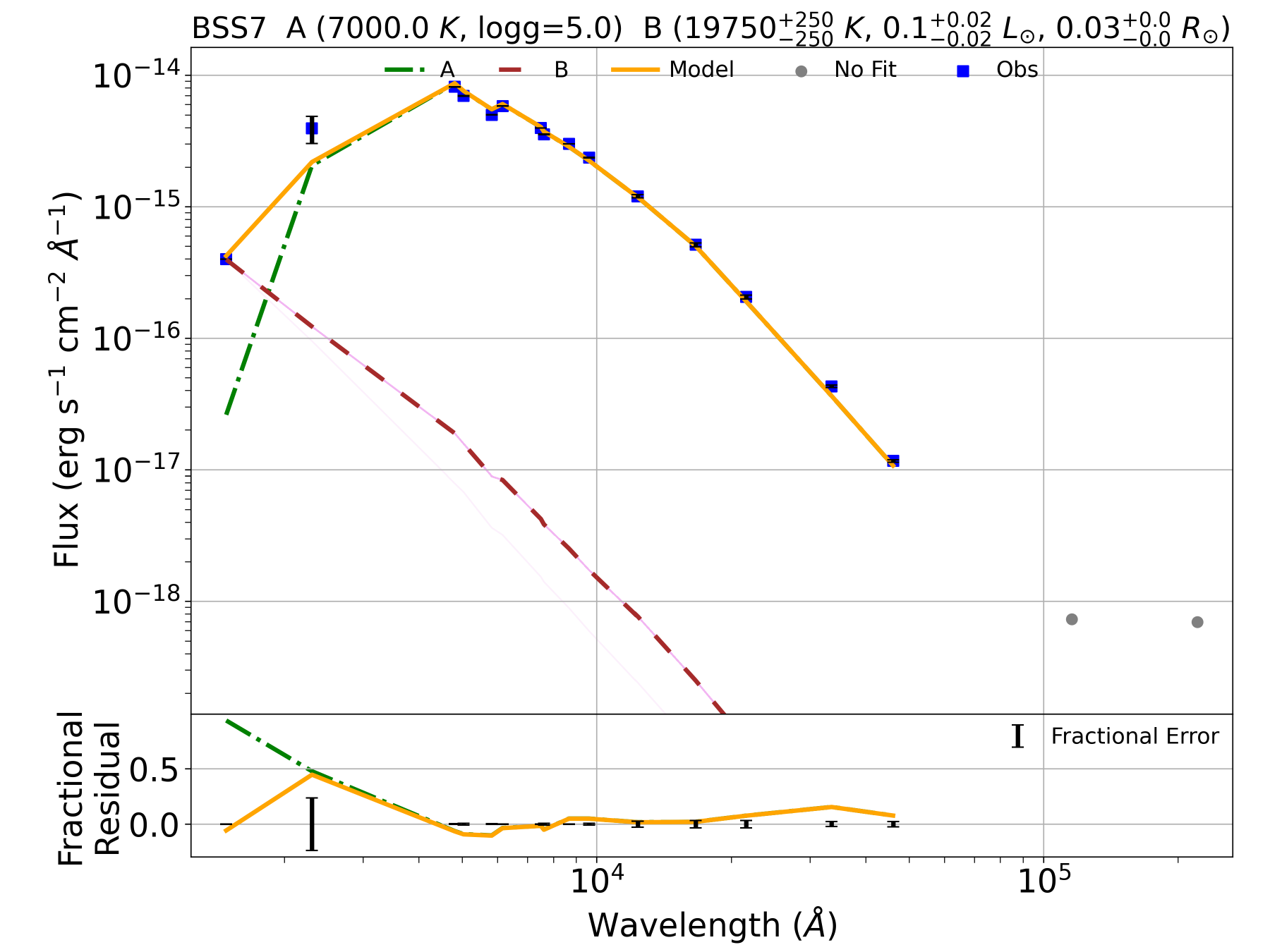}  
         \caption*{}  
     \end{subfigure}    
    \begin{subfigure}[b]{0.48\textwidth}
        \includegraphics[width=0.8\textwidth]{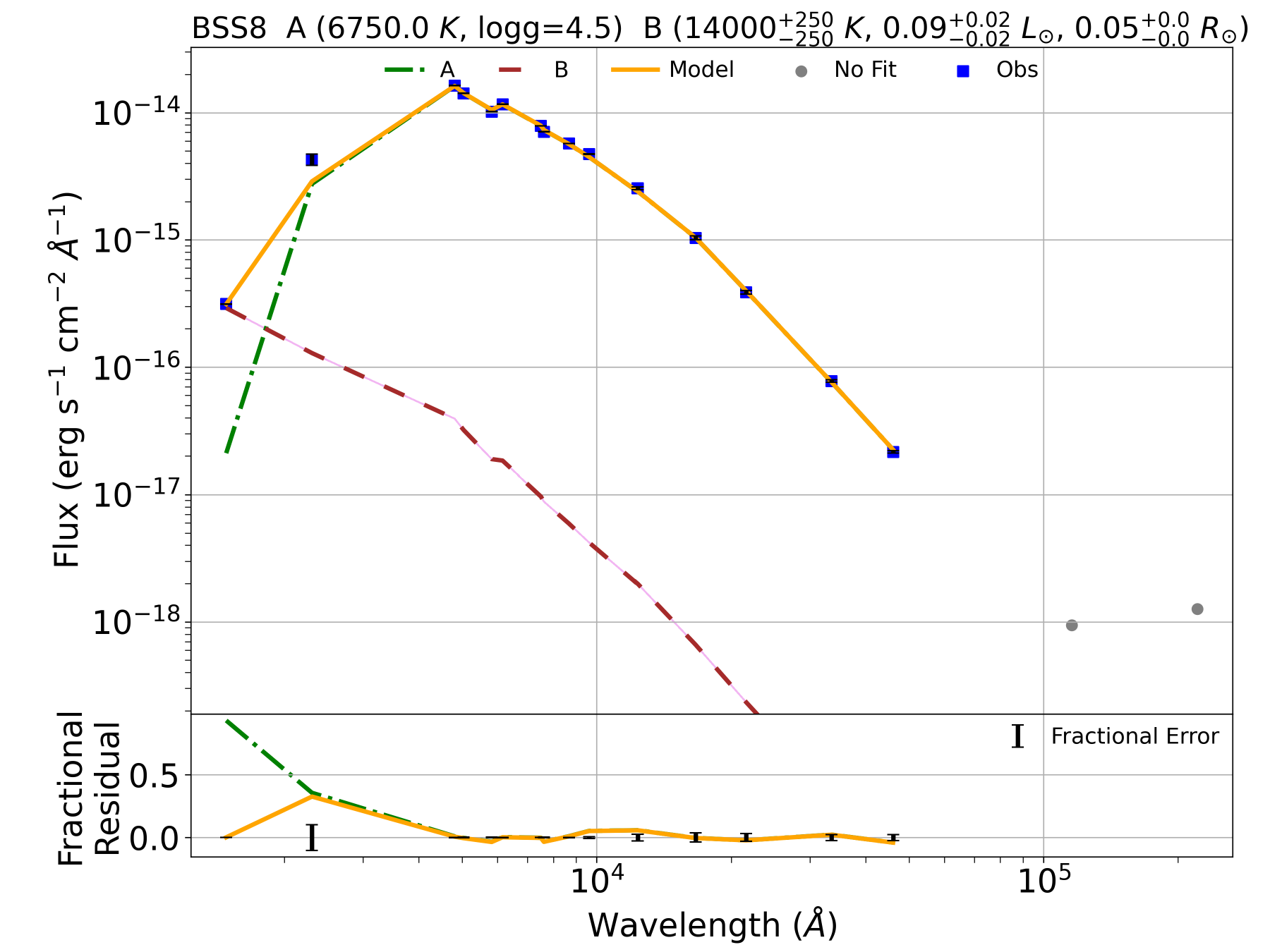}
        \caption*{}
    \end{subfigure}
    \quad
\caption{The binary-component SED of BSSs. The top panel displays the double component SED of each BSS, with blue data points representing extinction corrected flux values and black error bars showing flux errors. The green dashed line represents the cool (A) component fit, while the brown dashed line represents the hot (B) component fit, with pink curves representing iteration residuals. The composite fit is depicted as an orange curve, with grey data points denoting data points that were not included in the fits. The fractional residual for single (green) and composite (orange) fits is shown in the bottom panel. On the x-axis, black error bars represent fractional errors. The cool and hot component parameters derived from the SED, as well as the estimated errors, are listed at the top of the plots.}
\label{Fig.8}
\end{figure*}

In order to fit the binary component SEDs, we used a python code \textsc{Binary SED Fitting}\footnote{https://github.com/jikrant3/Binary SED Fitting}, which is based on $\chi^{2}_{r}$ minimization technique \citep{jadhav2021uocs}. We used the Koester model \citep{koester2010white} for this purpose, as this gives a temperature range of 5000 -- 80000 K and log g range of 6.5 -- 9.5, thus making it suitable to fit a compact hotter companion. The double component SEDs for all the BSSs are shown in Figure \ref{Fig.8}. The top panel for each BSSs shows the fitted SED and the bottom panel shows the residual for single and composite fit in each filter. We note that the residuals come out to be nearly zero on fitting the binary component SED since the excess in UV fluxes have reduced significantly. The parameters of all the BSSs are listed in Table \ref{Table2}, where the parameters of the cooler components are taken from VOSA, and that of the hotter companions are taken from the python code mentioned above. The errors in the parameters are determined by following the statistical approach mentioned in \cite{2021JApA...42...89J}.  The $\chi^{2}$ of the fits are found to be large even when the SED fits are visually good, mostly due to very small observational flux errors \citep{rebassa2021white}. We estimate modified reduced $\chi^{2}$, visual goodness of fit (vgf$_{b}$), determined by VOSA by forcing the observational errors to be at least 10$\%$ of the observed flux. The vgf$_{b}$ values $<$ 15 implies that the fits are good \citep{jimenez2018white,rebassa2021white}. 

\begin{table*}
\caption{The best-fit parameters of BSSs fitted with the double-component SEDs. For each of them, whether cooler (A) or hotter (B) companion in Column 2, luminosity, temperature, and radius in Columns 3 $-$ 5, the reduced $\chi^{2}_{r}$ values in Column 6 (the $\chi^{2}_{r}$ values of the single fits are given in the brackets), scaling factor in Column 7, number of data points used to fit the SEDs is given in Column 8, and the values of vgf$_{b}$ parameter in Column 9 (the vgf$_{b}$ values of the single fits are given in the brackets).}

%	\centering
	\adjustbox{max width=\textwidth}{
	\begin{tabular}{cccccccccccc}
		\hline
		\\
Name&Component&Luminosity&T$\mathrm{_{eff}}$&Radius&$\chi^{2}_{r}$&Scaling factor&N$_{fit}$&vgf$_{b}$\\
	    ~&~&[L$_\odot$]&[K]&[R$_\odot$]&&~&~&
        \\
		\hline 
		\\
BSS 1&A&26.97$\pm$0.03&6500$\pm$125&4.09$\pm$0.01&25.11 (211.7)&1.51E-21&15&0.55 (1.94)\\
&B&0.08$^{+0.02}_{-0.01}$&15000$\pm$125&0.04$\pm$0.0&-&&-&-\\
BSS 2&A&10.11$\pm$0.01&6750$\pm$125&2.35$\pm$0.00&267.90 (3647.01)&5.03E-22&15&0.24 (1.42)\\
&B&0.05$\pm$0.02&28000$\pm$1000&0.01$\pm$0.0&-&&-&-\\
BSS 3&A&11.00$\pm$0.01&7000$\pm$125&2.25$\pm$0.00&41.74 (111.4)&4.62E-22&15&0.15 (5.11)\\
&B&0.06$^{+0.02}_{-0.01}$&22000$\pm$1000&0.02$\pm$0.00&-&&-&-\\
BSS 5&A&12.27$\pm$0.01&7500$\pm$125&2.08$\pm$0.00&159.45 (439.1)&3.93E-22&15&0.25 (2.61)\\
&B&0.03$\pm$0.01&21000$\pm$1000&0.01$\pm$0.00&-&&-&-\\
BSS 7&A&8.72$\pm$0.01&7000$\pm$125&2.04$\pm$0.07&78.92 (376.08)&3.77E-22&15&0.57 (0.94)\\
&B&0.1$\pm$0.02&19750$\pm$250&0.03$\pm$0.0&-&-&-\\
BSS 8&A&16.87$\pm$0.02&6750$\pm$125&3.01$\pm$0.11&38.23 (302.1)&8.22E-22&15&0.12 (2.75)\\
&B&0.09$\pm$0.02&14000$\pm$125&0.05$\pm$0.00&-&&-&-\\
\hline
\label{Table2}
\end{tabular}
}
\end{table*}

\subsection{Binary modelling of the light curves}

Binary modelling is an excellent technique to obtain constraints on the stellar orbital parameters of a binary system. The additional constraints help us to understand the possible formation mechanism of the BSSs. In this study, we used PHysics Of Eclipsing BinariEs (PHOEBE) v0.32 \citep{prvsa2005computational} to model the two BSSs (BSS 1 and BSS 7) systems. This software package is an open-source eclipsing binary modelling code that theoretically calculates the stellar and orbital parameters of an eclipsing binary system using a synthetic light curve and a radial velocity curve. This is based on the Wilson-Devinney code \citep{wilson1971realization} and accepts an observed light curve and radial velocity measurements as inputs, as well as initial guesses for parameters such as effective temperature, surface gravity, inclination, semi-major axis, luminosity, and surface potential. 

The modelling of the two systems was performed in the following steps. Initially, we manually changed the parameters to obtain a close solution to the light curve. However, due to the unavailability of the radial velocity data for both systems, we used the Q-search method to determine the mass ratios. In the Q-search method, we performed multiple modelling runs for the same system with different values of the mass ratio. The value corresponding to the lowest sum of the squared deviations is taken as the mass ratio of the system. Once the mass ratio and an initial reasonable LC are obtained, we carry out iterative minimisation to find the best-fit model. The differential correction routine, available in PHOEBE, is performed repeatedly, where the next iteration is performed using the values of the previous cycle. This process is continued till the convergence of the cost function of the fitting. The parameters of BSS 1 and BSS 7 derived by fitting the binary model to the LCs using PHOEBE are listed in Table \ref{Table3}.

It is noteworthy that only two BSSs showed the variability signature in TESS, despite the fact that many may also be binary systems with hot companions and variables found in the literature. This could be due to several reasons. A binary system becomes an eclipsing binary only if the star is able to occult the light of the other, which is possible either if the system has high inclination or the radii of the companions are large. The BSSs in this cluster are found to have compact objects like white dwarfs as companions, which have a low probability of exhibiting eclipses. A similar system is shown in \cite{vernekar2023photometric}.
Furthermore, even if many more systems are eclipsing, their detection is limited by the amount of data available using TESS. Therefore, the BSSs system with longer periods cannot be probed using this method.

\subsection{Properties of Blue Straggler Stars}

On fitting the SED of the BSSs using the method described above, we found that the temperatures of the BSSs vary from 6500 K to 8000 K, which is consistent with the age of the cluster. Furthermore, the luminosities vary from 8.72 to 26.91 L$_{\odot}$, and radii vary from 2.04 R$_{\odot}$ to 4.09 R$_{\odot}$. Upon comparing temperatures derived from fitting the SED with the temperatures based on BP/RP spectra-based from \textit{Gaia} DR3 \citep{babusiaux2022gaia}, we found that the SED-based temperatures are within 400 K with the low-resolution spectra-based temperatures. We also find the rough estimation of masses of the BSSs by comparing it with the ZAMS of age 8.5 Gyr. The masses of the BSSs vary from 1.60 M$_{\odot}$ to 2.15 M$_{\odot}$. Since the turn-off mass of the cluster is 1.55 M$_{\odot}$, we suggest that the BSSs are likely to have gained $\sim$0.05 M$_{\odot}$ to 0.60 M$_{\odot}$ via MT/merger channel.

\begin{figure*}
\centering
\includegraphics[width=0.6\textwidth]{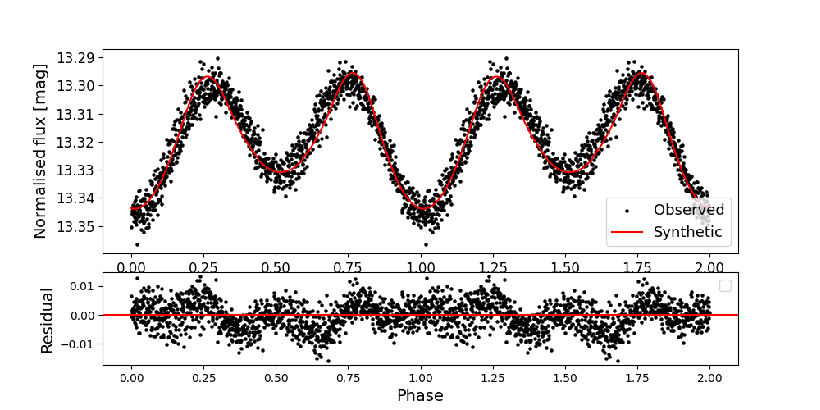}
\includegraphics[width=0.52\textwidth]{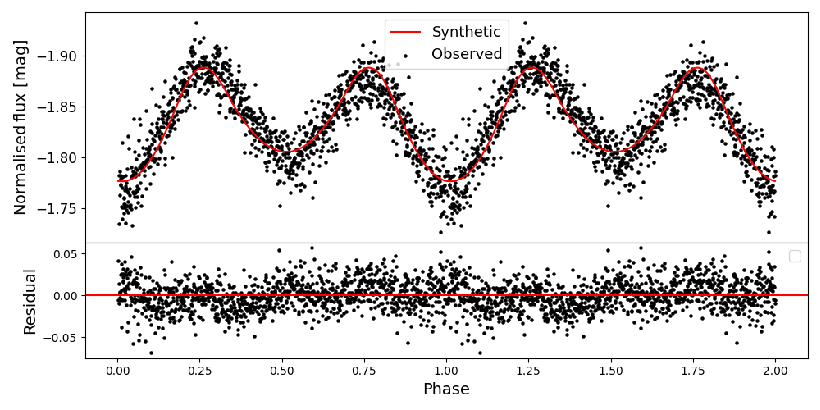}
\caption{The binary model fitted LCs of BSS 1 and BSS 7. The upper panels shows the LC, where the black dots represents the observed LC and the red line represents the synthetic LC. The lower panels shows the residual between observed and the synthetic LCs, which shows a good agreement which each other.}
\label{Fig.9}
\end{figure*}

\begin{figure*}
\centering
\includegraphics[width=0.4\textwidth]{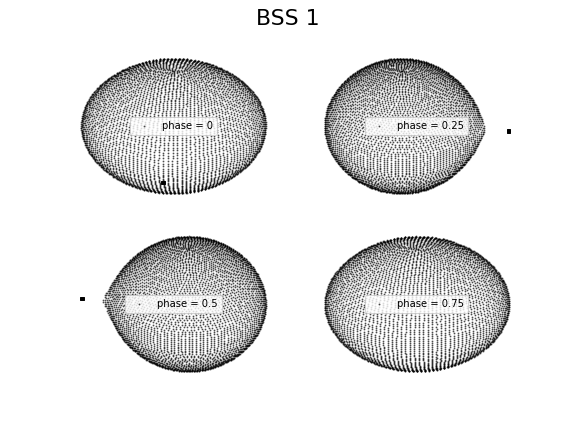}
\vrule
\includegraphics[width=0.4\textwidth]{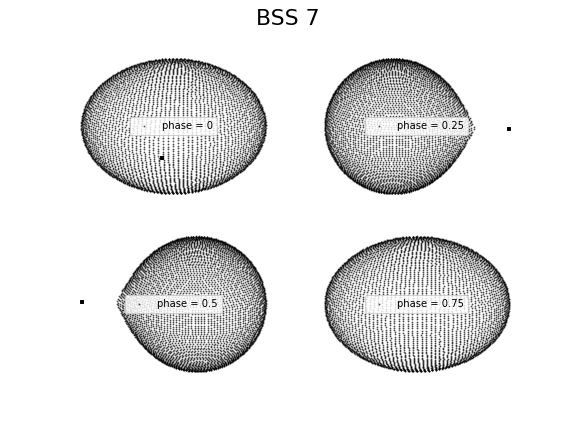}
\caption{The mesh plot showing the binary systems of BSS 1 in the left panel and of BSS 7 in the right panel shown in different phases. In each of them, the bigger object represents the BSS, whereas the smaller object represents the WD companion.}
\label{Fig.10}
\end{figure*}

We performed binary modelling of the LCs by fixing the temperatures of the BSSs generated from the SED. The binary LCs of BSS 1 and BSS 7 fitted with the model using PHOEBE are shown in Figure \ref{Fig.9}. The observed and synthetic LCs are in agreement with each other, which is well conveyed by the residuals shown in the lower panels. This implies that the temperatures determined by constructing the SED are much accurate. In the case of BSS 1, the radius derived using SED fitting is 4.09 $\pm$ 0.01 R$_{\odot}$ and that from binary modelling of LC is 3.55 $\pm$ 0.12 R$_{\odot}$, which are consistent with each other. However, in the case of BSS 7, the radius of the BSSs derived by constructing the SED is 2.04 $\pm$ 0.04 R$_{\odot}$ and that from the binary modelling of LC is 5.69 $\pm$ 0.29 R$_{\odot}$. The parameters of both the BSSs derived from fitting the LC are listed in Table \ref{Table3}. Figure \ref{Fig.10} shows the mesh plot of both the BSSs systems in different phases. The deformation of the blue straggler is indeed dominating the entire light curve with a small effect from eclipses. 

We also checked for the information on the variability of BSSs in \textit{Gaia} DR3 \citep{eyer2023gaia}, where BSS 1 and BSS 7 are reported as eclipsing binaries with the period of $\sim$1.07 and 0.58 days. On the other hand, BSS 3 and BSS 8 are found to be short time-scale ($<$ 0.5--1 d) MS-type oscillators (GDOR | DSCTU | SXPHE). However, for BSS 3 and BSS 8, we did not find any signatures of variability using TESS data.

\begin{table*}
\centering
\caption{Parameters of BSS 1 and BSS 7 obtained by fitting the binary model to the LCs using PHOEBE. For both of them, the starting time of the LC in Row 1, orbital periods in Row 2, the phase shift in Row 3, the mass ratio in Row 4, the inclination of the orbits in Row 5, eccentricity in Row 6, temperatures of BSSs (kept fixed) in Row 7, temperatures of the hotter companions in Row 8, the radius of BSSs in Row 9, and radius of the hotter companions in Row 10.}

\begin{tabular}{cccccccccccc}
\hline
\hline

Parameters&BSS 1&BSS 7 \\
\hline
\\

HJD0 (days)&2458738.18$\pm$0.015&2451738.319$\pm$0.015 \\
Period (days) &1.07066&0.5807	 \\
Phase shift&0.01456$\pm$0.014&0.0116$\pm$0.03 \\
Mass ratio&0.0257$\pm$0.0009&0.1142$\pm$0.0057 \\
Inclination&58.35$\pm$0.05&75.62$\pm$0.64 \\
Eccentricity&0.005$\pm$0.002&0.001$\pm$0.002 \\
T$_{eff}$ (cooler)(K)&6000&7000 \\
T$_{eff}$ (hotter) (K)&15000$\pm$4500&20000$\pm$7000 \\
Radius (cooler) (R$_{\odot}$)&3.55$\pm$0.12&5.69$\pm$0.29 \\
Radius (hotter) (R$_{\odot}$)&0.038$\pm$0.011&0.037$\pm$0.013\\

\hline
\label{Table3}
\end{tabular}
\end{table*}

\subsection{Properties of Hot Companions}

\begin{figure*}
\centering
\includegraphics[width=0.7\textwidth]{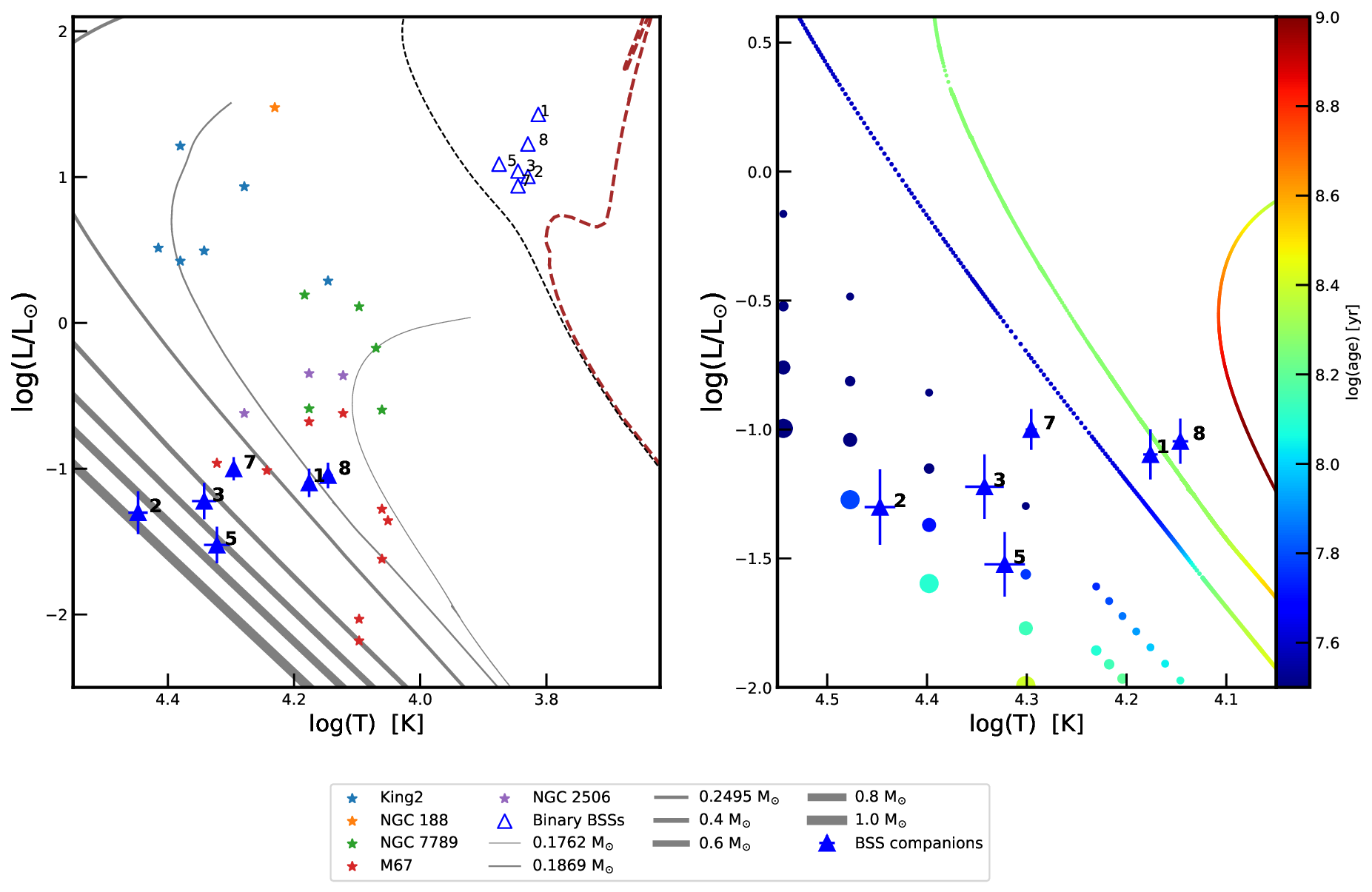}
\caption{The left panel contains a PARSEC isochrone with an age of 4.0 Gyr and distance of 2368 pc along with a zero-age main sequence (ZAMS), which is constructed by stitching ZAMS of ages 8.0 Gyr, 8.5 Gyr, and 9.0 Gyr. We have shown the cooler and hotter companions of NGC 7142 BSSs, along with the hot BSSs companions of other OCs. The WD cooling curves of different masses which are taken from \citet{althaus2013new} in the case of ELM WDs, \citet{panei2007full} in the case of LM WDs, and \citet{tremblay2009spectroscopic} in the case of normal to high mass WDs are also shown. The right panel depicts the hot companions of BSSs lying on the WD cooling curves of different masses, indicating their approximate cooling ages. 
}
\label{Fig.11}
\end{figure*}

In order to gain a complete understanding of how the BSSs have formed, examining the nature of their hotter companions is crucial. In order to achieve this, we constructed the Hertzsprung-Russell (H-R) diagram as shown in Figure \ref{Fig.11}. The locations of the hotter companions of BSS 1 and BSS 8 in the left panel of this figure indicate that they are ELM WDs \citep{althaus2013new} of mass 0.18 M$_{\odot}$. The stellar remnants known as ELM WDs do not ignite helium in their cores \citep{brown2011binary}. The hotter companion of BSS 7, on the other hand, lies on the WD cooling curve with a mass of 0.20 M$_{\odot}$ \citep{panei2007full}, indicating that it is an LM WD. The discovery of ELM/LM WD companions supports the Case-A/Case-B MT formation mechanism of BSS because ELM WDs and LM WDs cannot form from a single star evolution within Hubble time \citep{brown2011binary}. This implies that these WDs must have lost mass during their evolution. Furthermore, the hotter companions of BSS 3 and BSS 5 are likely to be normal-mass WDs, suggesting Case-C MT as the possible formation channel. The hotter companion of BSS 2 is lying on the WD cooling curve of a high-mass WD, suggesting merger in a hierarchical triple system as the formation mechanism of this BSS. Moreover, the radii of the hotter companions of BSS 1 and BSS 7 derived from fitting the SED and fitting the LCs are in good agreement. In the case of BSS 1, the SED-based radius of the hotter companion is 0.04 $\pm$ 0.00 R$_{\odot}$, whereas the LC-based radius is 0.038 $\pm$ 0.011 R$_{\odot}$. Similarly, in the case of BSS 7, the SED-based radius is 0.03 $\pm$ 0.00 R$_{\odot}$ and that derived from LC fitting is 0.037 $\pm$ 0.013 R$_{\odot}$. The reason for such a large error in the parameters of the secondary companion is that the whole LC is dominated by the luminosity variation of the deformed primary. Only a small portion of the light variation arises due to eclipse. Since the parameters such as T$_{eff}$ and radius are measured from the width and depth of the eclipses, the ability to obtain good constraints is not possible, leading to large errors. 

It is evident from the right panel of Figure \ref{Fig.11}, that the log age of hot companions of BSS 1 and BSS 8 are $\sim$ 8.3, implying that these BSSs are formed via MT almost 0.2 Gyr ago. Moreover, the log age of BSS 3 and BSS 7 WDs are $\sim$7.6, which suggests that these objects were formed 30 Myr ago. On the other hand, the hot companions of BSS 2 and BSS 5 are lying on the WD cooling curves of log age $\sim$7.9, which suggests that these BSSs were formed $\sim$ 70 Myr ago.

This is the first-ever OC studied using \textit{AstroSat}/ UVIT data that has hotter companions of BSSs with masses varying from 0.18 M$_{\odot}$ to 1.0 M$_{\odot}$. The most comprehensive study of a BSSs population in an OC is NGC 188, where 80$\%$ of the BSS are found to be SB1s with typical orbital periods, P $\sim$ 1000 days\citep{mathieu2009binary,gosnell2015implications}. Additionally, \cite{geller2011mass} found that the statistical mass distribution of the companions to these BSSs peaks around 0.5 M$_{\odot}$, strongly indicating the presence of WD secondary stars. Subsequent to that, Hubble Space Telescope UV photometry further confirmed that seven out of the 20 BSSs in NGC 188 possess hot WD companions with an age less than 400 Myr. This suggests recent MT events from RGB or AGB companions \citep{gosnell2014detection,gosnell2015implications}. Various studies based on UVIT/\textit{AstroSat} mentioned in Section \ref{Section 1}, including this work, also suggest MT as the predominant mechanism for the formation of BSSs in OCs. Even in the case of GCs, the absence of a correlation between cluster density and BSSs frequency implies that internal binary evolution, including MT, is the primary formation pathway, as opposed to dynamical collisions \citep{knigge2009binary}. Furthermore, the presence of a significant population of BSs and other AFG-type post-MT binaries in the field suggests that MT is likely a prevalent formation mechanism across various environments \citep{carney2001survey,murphy2018finding, escorza2019barium, panthi2023field}. Collectively, these populations provide valuable insights into the outcomes of post-MT events in solar-type binaries. However, detailed investigations of BSSs populations, particularly across diverse populations encompassing different cluster ages, densities, and metallicities, remain highly relevant.

\section{Summary and conclusions} \label{Section 5}

The work presented in this paper can be summarized as follows: 

\begin{enumerate}

\item We use a machine learning-based algorithm, ML-MOC, on \textit{Gaia} DR3 data, to identify 546 cluster members in NGC 7142, including 10 BSSs. We analyse these BSSs using \textit{AstroSat}/UVIT data and complementary data from other archival data in a single FUV filter, F148W. 

\item We construct the multi-wavelength SED of seven out of ten BSSs. Three BSSs had a neighbouring source within 3\arcsec, and hence their SEDs are not reliable. It is interesting to note that all the seven BSSs showed excess in UV wavelengths and, hence, likely to have the hotter companions associated with them. Out of these seven BSSs, we constructed the binary-component SED of six of them. In the case of the remaining one BSSs, the hotter companion was fitting the highest possible temperature; hence, we show only the single-component fit of this star. This work reports the properties of these six BSSs and their hotter companions. 

\item An ELM WD with mass $\sim$0.18 M$_{\odot}$, temperature $\sim$ 15000 K, luminosity $\sim$ 0.08 L$_{\odot}$, and radius $\sim$0.04 R$_{\odot}$ is found to be the hotter companion of BSS 1. Moreover, we discover a LM WD of mass $\sim$ 0.30 M$_{\odot}$ with T$_{eff}$ $\sim$19750 K, L $\sim$0.10 L$_{\odot}$, and R $\sim$0.03 R$_{\odot}$ as a companion of BSS 7. The discovery of ELM and LM WDs suggests that these BSS are formed via Case-A/Case-B MT. We construct the TESS LCs of 10 BSSs candidates of this cluster, out of which BSS 1 and BSS 7 showed the signatures of variability. The temperatures and radii of BSS 1 and BSS 7 as well as their hotter companions obtained by constructing the SED and fitting the binary models to the TESS LCs, are in good agreement with each other. We confirm that BSS 1 and BSS7 are indeed eclipsing binaries using the TESS LCs along with the estimation of their orbital parameters.

\item The second ELM WD of mass $\sim$ 0.18 M$_{\odot}$, temperature $\sim$ 15000 K, luminosity $\sim$ 0.09 L$_{\odot}$, and radius $\sim$ 0.10 R$_{\odot}$ is found as a companion of BSS 8. Furthermore, a normal-mass WD with a mass of $\sim$0.5 M$_{\odot}$, temperature of 22000 K, L $\sim$ 0.06 L$_{\odot}$, and R $\sim$0.02 R$_{\odot}$ is found to be the companion of BSS 3. Another BSS, BSS 5, has a normal-mass WD of mass $\sim$0.6 M$_{\odot}$ with the estimated temperature of 21000 K, luminosity of $\sim$0.03 L$_{\odot}$ and a radius of 0.01R$_{\odot}$, as a hot companion. A high-mass WD of mass $\sim$1.0 M$_{\odot}$ is discovered as the hot companion of BSS 2, with an estimated temperature of 28000 K, luminosity $\sim$0.05 L$_{\odot}$, and radius $\sim$0.01 R$_{\odot}$. These normal/high-mass WDs point towards the Case-C MT/merger as possible formation channels. 

\item There is even a possibility that the BSS 6, which was not successfully fitted with a second companion may as well be a binary as it showed UV excess. We confirm that out of 10 BSSs, at least 6 have WDs as the hot companions, and at least 3 BSSs are formed via the Case-A/Case-B MT channel. Moreover, the remaining three are formed via either Case-C MT or merger formation channels. 

\end{enumerate}

\section*{Acknowledgements}
We thank the anonymous referee for the valuable insights and constructive feedback, which greatly contributed to the enhancement of the manuscript. AS thanks for the support of the SERB power fellowship. VJ thanks the Alexander von Humboldt Foundation for their support. This work uses the data from UVIT onboard the \textit{AstroSat} mission of the Indian Space Research Organisation (ISRO). UVIT is a collaborative project between various institutes, including the Indian Institute of Astrophysics (IIA), Bengaluru, The Indian-University Centre for Astronomy and Astrophysics (IUCAA), Pune, Tata Institute of Fundamental Research (TIFR), Mumbai, several centres of Indian Space Research Organisation (ISRO), and Canadian Space Agency (CSA). This publication also makes use of VOSA, developed under the Spanish Virtual Observatory project supported by the Spanish MINECO through grant AyA2017-84089. 

\section*{Data availability}

The data underlying this article are publicly available at \url{https://astrobrowse.issdc.gov.in/astro_archive/archive/Home.jsp} The derived data generated in this research will be shared on reasonable request to the corresponding author.

\bibliography{references}
\bibliographystyle{MNRAS}

\bsp	
\label{lastpage}
\end{document}